# The Role of the Magnetic Anisotropy in Atomic-Spin Sensing of 1D Molecular Chains

*Christian Wäckerlin,[a,b,*] Aleš Cahlík,[a] Joseba Goikoetxea,[c] Oleksandr Stesovych,[a] Daria Medvedeva,[d] Jesús Redondo,[a] Martin Švec,[a] Bernard Delley,[e] Martin Ondráček,[a] Andres Pinar,[a] Maria Blanco-Rey,[f,g] Jindrich Kolorenc,[d,*] Andres Arnau,[c,f,g,*] Pavel Jelínek[a,g*]*

[a] Institute of Physics, Czech Academy of Sciences, Cukrovarnická 10, 16200 Prague, Czech Republic, e-mail: jelinekp@fzu.cz

[b] Surface Science and Coating Technologies, Empa, Swiss Federal Laboratories for Materials Science and Technology, Überlandstrasse 129, 8600 Dübendorf, Switzerland, e-mail: christian.waeckerlin@empa.ch

[c] Centro de Física de Materiales CFM/MPC (CSIC-UPV/EHU), Paseo Manuel de Lardizábal 5, 20018 Donostia-San Sebastián, Spain, e-mail: andres.arnau@ehu.eus

[d] Institute of Physics, Czech Academy of Sciences, Na Slovance 2, 18221 Prague, Czech Republic, e-mail: kolorenc@fzu.cz

[e] Condensed Matter Theory, Paul Scherrer Institut, CH-5232 Villigen, Switzerland

[f] Departamento de Polímeros y Materiales Avanzados: Física, Química y Tecnología, Facultad de Química UPV/EHU, Apartado 1072, 20080 Donostia-San Sebastián, Spain

[g] Donostia International Physics Center (DIPC), Paseo Manuel de Lardizábal 4, 20018 Donostia-San Sebastián, Spain

ABSTRACT

One-dimensional metal-organic chains often possess a complex magnetic structure susceptible to be modified by a alteration of their chemical composition. The possibility to tune their magnetic properties provides an interesting playground to explore quasiparticle interactions in low-dimensional systems. Despite the great effort invested so far, a detailed understanding of the interactions governing the electronic and magnetic properties of the low-dimensional systems is still incomplete. One of the reasons is the limited ability to characterize their magnetic properties at the atomic scale. Here, we provide a comprehensive study of the magnetic properties of metal-organic one-dimensional (1D) coordination polymers consisting of 2,5-diamino-1,4-benzoquinonediimine ligands coordinated with Co or Cr atoms synthesized in ultra-high vacuum conditions on a Au(111) surface. A combination of an integral X-ray spectroscopy with local-probe inelastic electron tunneling spectroscopy corroborated by multiplet analysis, density functional theory, and inelastic electron tunneling simulations enable us to obtain essential information about their magnetic structure, including the spin magnitude and orientation at the magnetic atoms, as well as the magnetic anisotropy.

One-dimensional magnetic metal-organic coordination chains are a type of molecular nanomagnets that can be considered as candidates for building single-chain magnets under certain conditions.[1] Nowadays, a complete understanding of the relation between the chemical composition of metal-organic systems (defined by the type of organic ligand and magnetic atom constituents)[2] and relevant magnetic parameters (*e.g.* the strength and kind of the intrachain exchange coupling or the magnetic anisotropy energy) is still missing. Knowledge of these quantities is essential to determine the critical temperature, at which magnetic order prevails, or the characterization of the magnon excitation spectrum. Furthermore, these systems have potential applications in the fabrication of nanodevices for spin sensing, spintronics, quantum computing based on the spin degree of freedom or high-density information storage.[3]

Additionally, magnetism in low-dimensional (n-D) systems is also interesting from the fundamental point of view. For example, as Mermin-Wagner theorem[4] forbids magnetic order in isotropic low-$n$ ($n \leq 2$) systems, the observation of antiferromagnetic (AFM) or ferromagnetic (FM) order in these systems calls for an explanation. However, the study of the low-dimensional magnetic systems, often exhibiting strong electron correlations, is not currently trivial due to the limits of available experimental and theoretical tools. One option is to use ensemble techniques, such as X-ray absorption spectroscopy (XAS), X-ray linear dichroism (XLD) and X-ray magnetic circular dichroism (XMCD). These are powerful techniques but the accurate determination of magnetic properties from the acquired spectra is possible only with the help of complex theoretical many-body calculations.

Another option is based on the use of inelastic electron tunneling spectroscopy (IETS), which permits to locally probe both vibrational[5–14] and spin excitations[15–23]. Unfortunately, due to the strong interaction with itinerant electrons in the metallic electrodes, the inelastic tunneling channels of most magnetic adsorbates are extremely weak. For a long time, this has severely hampered the general applicability of STM-IETS to study single atom spin excitations. A novel approach based on IETS measurements consists of using nickelocene (NiCp$_2$, Figure 1a) decorated tips because they allow detecting magnetic exchange interactions with spins of individual atoms at surfaces *via* the spin excitations of the NiCp$_2$ probe and their mutual coupling.[24,25]

Part of us have recently demonstrated synthesis of long (> 100 nm), isostructural metal-organic 1D coordination polymers formed by reaction of transition metal (TM) atoms with 2,5-diamino-1,4-benzoquinonediimine (QDI), see Figure 1b.[26] Here, we provide a thorough analysis of the magnetic structure of TM-QDI chains (with TM = Cr, Co) grown on the metallic Au(111) surface by using multiple experimental and theoretical techniques. In particular, we use XAS, XLD and XMCD combined with multiplet calculations to determine the 3d electronic structure and magnetic properties of CrQDI and CoQDI coordination polymers on Au(111). This analysis is complemented by density functional theory (DFT) calculations using GGA+U and hybrid exchange correlation functionals. The obtained information about the 3d orbital occupation and ligand field is used to rationalize the IETS spectra recorded on the two systems with a NiCp$_2$ functionalized tip using an original methodology for IETS simulations.

By a combination of these experimental and theoretical methods, we can unravel the magnetic properties of the metal-organic chains. Specifically, we can determine: i) the spin multiplicity, *i.e.* decide between high-spin (HS) states *vs.* low-spin (LS) states, ii) the magnetic anisotropy, *i.e.* the easy axis of magnetization and the magnitude of the anisotropy, as well as iii) why can NiCp$_2$-IETS exhibit no exchange splitting despite probing a magnetic center.

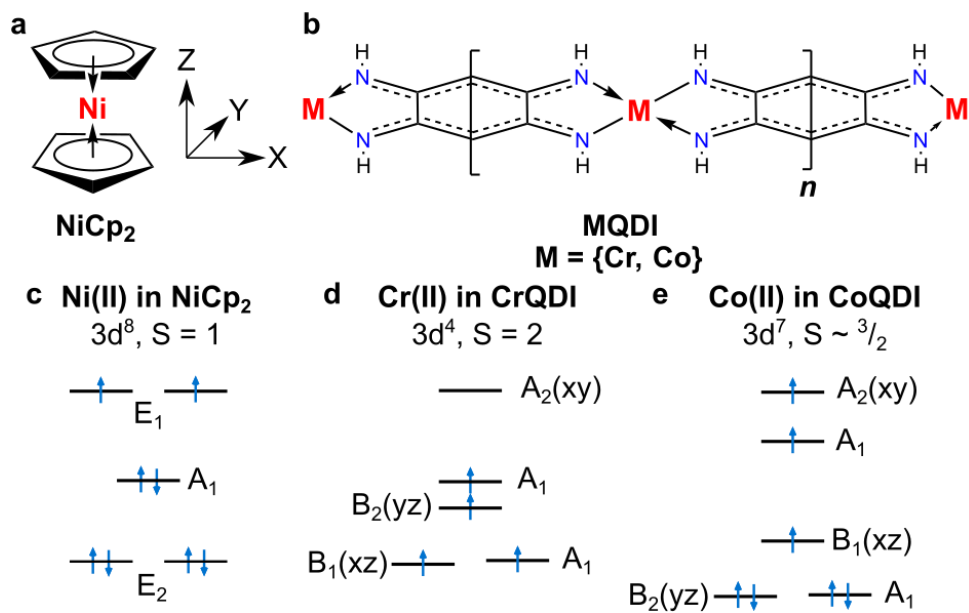

**Figure 1**: Scheme of (a) the probe molecule nickelocene (NiCp$_2$) and (b) the chromium and cobalt based metal-organic diamino-benzoquinonediimine (MQDI) coordination polymers on Au(111). NiCp$_2$ is attached to the STM tip by its top Cp ring and brought over the MQDI wires for IETS experiments. Also shown are 3d manifolds of c) NiCp$_2$, d) CrQDI and e) CoQDI. The high-spin $S$ = 2 and $S \sim 3/2$ configurations of CrQDI and CoQDI are determined by XAS/XLD (*vide infra*).

**Results and Discussion**

**X-ray Spectroscopy Experiments and Multiplet Modeling**

Figure 2 shows Co $L_{3,2}$ and Cr $L_{3,2}$ XAS, X-ray linear dichroism and X-ray magnetic circular dichroism. In combination with effective point-charge atomic multiplet calculations using the MultiX software,[27] the data permit the extraction of the electronic ground states and magnetization of the Co and Cr ions in CoQDI and CrQDI. The XAS/XLD spectra (Figure 2a,d), were calculated by optimizing the ligand field input using least squares fits. The corresponding details are described in the methods section.

The magnetic properties determined by XAS/XLD combined with the optimized multiplet model are reported in Table 1. The X,Y,Z directions are defined in Figure 1. They refer to the in-plane direction along the wire X, the in-plane direction perpendicular to the wire Y, and the out-of-plane direction Z. The parameters of the multiplet models are reported in Tables S1 and S2.

The multiplet model shows that Co and Cr ions in these systems have 3d occupations expected for the formal M(II) ions bridging two QDI ligands,[2] *i.e.,* $3d^7$ and $3d^4$, and that both ions are in their high spin states, *i.e.*, Co is in its $S \sim 3/2$ quadruplet state with the easy axis of magnetization along the Y direction and Cr is in its $S = 4/2$ quintet state with the easy axis of magnetization along the Z direction. Furthermore, Co exhibits a large orbital moment and, hence, a large zero-field splitting (ZFS) of several meV, while Cr in CrQDI exhibits a quenched orbital moment ($L_Z = 0.05$) and thus a much lower ZFS.

**Table 1. Magnetic properties of CoQDI and CrQDI determined by fitting the XAS/XLD data with multiX**

| Metal atom | 3d Occupation | Spin state | $m_s$ ($\mu_B$) | $m_l$ ($\mu_B$) | Easy-axis | ZFS (meV) |
|---|---|---|---|---|---|---|
| Co | $3d^7$ | $1.47 \cong 3/2$ (HS) | 2.94 | 1.79 | Y | 37* |
| Cr | $3d^4$ | 4/2 (HS) | 4 | 0.05 | Z | 0.7 |

* significant uncertainty

The multiplet model considers the metal Co or Cr ions to be purely paramagnetic, *i.e.* without any magnetic coupling between Co or Cr neighbors in the 1D chain and without spin quenching by Kondo interactions with the Au substrate. Under these assumptions, the model predicts large XMCD signals for Co in grazing incidence and for Cr in both grazing and normal incidence (Figure 2, dashed lines). Experimentally, no Co XMCD signal is observed, despite the low temperature (3 K) and the high magnetic field (6.8 T). Indeed, XMCD signal is detected only for Cr, its magnitude being significantly smaller than in the case of the multiplet simulations. The absent (Co) and weak (Cr) XMCD signals suggest the presence of antiferromagnetic (AFM) interactions in both chains. For CrQDI, the XMCD magnetization curves show higher magnetization in normal incidence than in grazing incidence (Figure S1), consistent with the out-of-plane easy axis of magnetization predicted by the multiplet simulations.

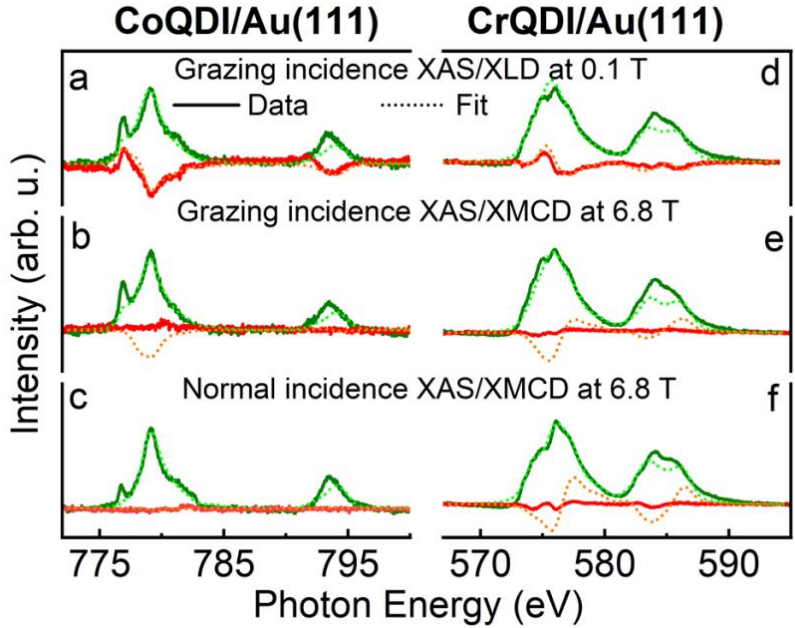

**Figure 2.** Experimental and simulated Co and Cr $L_{3,2}$ XAS, XLD and XMCD spectra of CoQDI (a-c) and CrQDI (d-f) on Au(111) recorded at 3 K. The XAS/XLD is used to fit the ligand field parameters within the multiplet code multiX. The XAS/XMCD signals are then simulated based on these parameters. The data are shown as solid lines (olive: XAS, red: XLD and XMCD) and the fitted/simulated spectra are shown as dotted lines (light green: XAS, orange: XLD and XMCD). The multiplet model of CoQDI gives a $3d^7$ ground state with high-spin state ($S \cong 3/2$) and a strong easy axis of magnetization in the in-plane direction perpendicular to the chain. For CrQDI, the corresponding model with the relevant parameters from the fit gives a $3d^4$ high-spin ($S = 2$) ground state with a weak easy axis of magnetization in the out-of-plane direction. Note that the multiplet model considers the metal atoms as paramagnets and, hence, it predicts large XMCD signals. The much lower experimental XMCD signal of CrQDI and the quenched XMCD signal of CoQDI can be explained by the presence of antiferromagnetic interactions (see below).

**Density Functional Theory**

To provide a better insight into the structural, electronic and magnetic properties of the CoQDI and CrQDI 1D metal-organic chains, we have carried out total energy DFT calculations of the free-standing polymers using GGA+U[28,29], as well as hybrid PBE0[30] exchange-correlation functionals. The values obtained from GGA+U method for various magnetic quantities of CoQDI and CrQDI chains are given in Table 2.

Interestingly, our GGA+U calculations with $U = 5$ eV yield a high-spin (HS) ground state for CoQDI with spin magnetic moment $m_S = 2.73$ $\mu_B$, consistent with the $S \cong 3/2$ quadruplet found in the multiplet analysis of XAS data (see the previous section). These calculations also show that a metastable low-spin (LS) state exists with $m_S = 1.10$ $\mu_B$, *i.e.*, compatible with an $S = 1/2$ doublet, at 88 meV from the ground state. The Co−N bond length in the LS state is 0.16 Å shorter than in the HS state. This sensitivity to the metal-organic bonding points to a subtle dependence of the Co multiplets on the screening of the intra-atomic Coulomb interactions in the Co atom by the QDI organic ligand. Therefore, we have explored the spin/geometry configurations space using DFT calculations within the GGA+U approach by varying the Coulomb parameter U and the 1D chain lattice constant. As shown in Figure 3, we find that between $U = 4$ and 5 eV there is a spin crossover from the LS state ($m_S = 1.20$ $\mu_B$) to the HS state ($m_S = 2.69$ $\mu_B$), alongside an extension of the Co−N bond length that amounts to 0.16 Å (see Table 2). The Co(d) spin-minority orbital configurations in the LS and HS ground states at $U = 4$ and 5 eV are $(d_{z^2})^1 (d_{x^2-y^2})^1 (d_{xz})^{0.75} (d_{xy})^{0.25}$ and $(d_{z^2})^1 (d_{x^2-y^2})^1$, respectively. The band structure and densities of states (PDOS) projected onto the Co(d) and N(p) orbitals of the $U = 5$ eV HS and the $U = 4$ eV LS ground states, together with the HOMO and LUMO in the spin majority and minority channels, are shown in Figures S3 and S4. The hybridization between bands with $d_{z^2}$ and $d_{x^2-y^2}$ character indicates that they stem from a sp³d² hybrid atomic orbital, *i.e.*, there is a substantial intra-atomic s−d charge transfer. Similar trends of a transition from LS to HS state are also predicted by hybrid PBE0 calculations, when the contribution of the exact-exchange component is increased with respect to default value, for details see Figure S8 to S11 and Table S7.

In the CrQDI case, DFT calculations using GGA+U with $U = 5$ eV, identify a unique ground state with lattice constant 8.14 Å, the corresponding N−Cr bond length being 2.08 Å, and with Cr atoms in a magnetic state $S = 2$ with an electronic configuration $(d_{yz})^1 (d_{z^2})^1 (d_{xz})^1 (d_{x^2-y^2})^1$, which translates into a spin magnetic moment $m_S = 3.71$ $\mu_B$. Calculations initialized with other occupancies revert to this ground state. The corresponding band structure and PDOS are shown in Figure S4. The HS configuration described above is also obtained for other $U$ values, as well as with the hybrid functional calculations using PBE0 (see Figure S6 and S7 and Table S7).

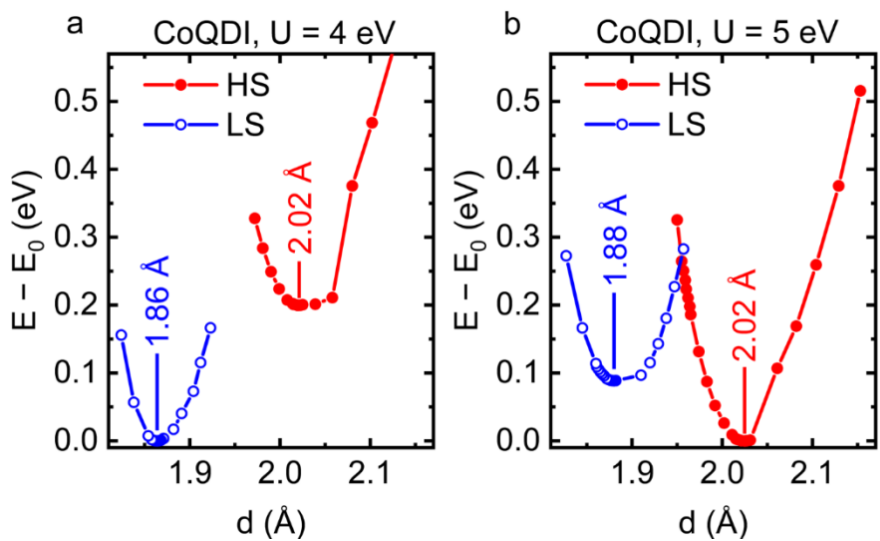

**Figure 3.** Dependence of the total energy on the N−Co bond-length (d) for low-spin (LS) and high-spin (HS) states of CoQDI for two different $U$ values: $U = 4$ eV (a) and $U = 5$ eV (b). The calculations show a spin crossover from the LS to HS ground state for $U = 4$ eV vs. $U = 5$ eV. The N−Co bond lengths of the local minima are shown. They differ by 0.16 Å between the true LS and HS ground states. The energies are referred to the respective ground states.

Calculations done with the GGA+U functional using a doubled unit cell to obtain the exchange coupling constant ($J$) yield weak antiferromagnetic coupling for CoQDI in the HS state, with an exchange coupling constant $J = -2.95$ meV, and weak ferromagnetism for the LS state with an exchange coupling constant $J = 2.04$ meV. These values are obtained from the total energy difference of two broken-symmetry solutions, corresponding to parallel (FM) and antiparallel (AFM) alignment of the spin magnetic moments on TM centers. The ground state of CrQDI is found to be AFM with an exchange coupling constant $J = -4.42$ meV, increasing in magnitude as U decreases (see SI). Importantly, the magnetic order is robust against geometric changes for both CoQDI and CrQDI chains, as checked by low-energy vibrational mode excitations of the chain (Figure S5). Moreover, a spin spiral analysis shows that the exchange with the second nearest neighbors (both for Co and Cr chains) is FM. Beyond second neighbors, exchange interactions significantly drop by one order of magnitude, meaning that there are no frustrated interactions in the coupling. Furthermore, when spin-orbit effects are included in the GGA+U calculations, we find a negligible anisotropic exchange. In other words, the dependence of the exchange coupling constant $J$ on the orientation of the Co or Cr spins is negligible (Figure S3).

**Table 2: Magnetic properties of CrQDI and CoQDI calculated by GGA+U.** $J$, $m_s$, $m_l$ and $E_{MAE}$ are the exchange coupling strength, the spin and orbital magnetic moments and the magnetocrystalline anisotropy energy. $\Delta E = E_{AFM} - E_{FM}$.

| Metal atom | Co−N bond-length (Å) | ΔE (meV) | $J$ (meV)* | $m_s$ (μB) | $m_l$ (μB) | $E_{MAE}$ (meV) | Easy axis |
|---|---|---|---|---|---|---|---|
| Co (LS, $U$ = 4 eV) | 1.86 | 1.48 | 2.05 (FM) | 1.20 | 0.200 | −0.76 (X−Z) | X |
| Co (HS, $U$ = 5 eV) | 2.02 | −11.07 | −2.97 (AFM) | 2.73 | 0.207 | −1.22 (Y−Z) | Y |
| Cr ($U$ = 5 eV) | 2.08 | −30.43 | −4.42 (AFM) | 3.71 | −0.038 | 0.54 (X−Z) | Z |

* $J = (E_{AFM} - E_{FM})/(2 \times S^2)$ with $S = m_s/2$

According to Goodenough-Kanamori (G-K) rule,[31,32] AFM super-exchange between spin magnetic moments localized at the 3d magnetic atoms appears when singly occupied 3d orbitals overlap with molecular orbitals of the ligand molecule. This is the case of Co(HS) and Cr metal-organic chains, which have hybrid bands with $d_{xz}$ or $d_{xy}$ character but not the case of Co(LS). As an example, to illustrate the G-K rule for the HS state of CoQDI, we have selected two singly occupied hybrid bands with $d_{xz}$ and $d_{xy}$ character that are coupled to molecular orbitals of the ligand with π (see Figure 4 left panel) and σ (Figure 4 right panel) character, respectively. Therefore, they both fulfill the criterion of the G-K rule to explain the tendency to have AFM (super) exchange coupling between the Co spins. Another interesting observation is that, among the three systems considered, the two insulating systems, CoQDI(HS) and CrQDI, correspond to AFM coupling and only the metallic system, CoQDI (LS), corresponds to FM coupling, following the previously observed trend that metallic hybrid bands tend to favor FM coupling between spins in metal-organic coordination networks.[33]

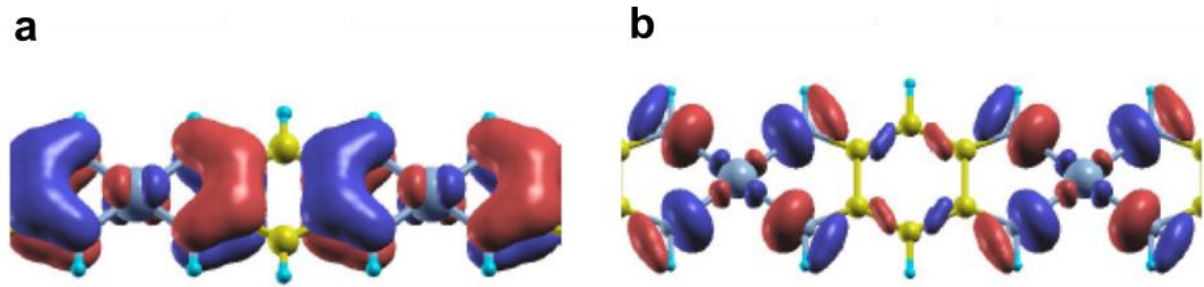

**Figure 4** Visualization of two orbitals corresponding to the CoQDI metal-organic chain showing singly occupied bands formed by the hybridization of Co($d_{xz}$) or Co($d_{xy}$) orbitals with π (a) or σ (b) molecular orbitals.

Finally, we present the results of our DFT self-consistent calculations including spin-orbit coupling to obtain the magnetic anisotropy energy (MAE) from the total energy difference of two different magnetization directions in the same system geometry. The results show that the easy magnetization direction lies in plane and is perpendicular to the chain for CoQDI HS, while for CrQDI the easy magnetization axis is out of plane, in agreement with the multiplet model fitted to the XAS/XLD data. For CoQDI HS, the magnetic anisotropy energy provided by the GGA+U calculations is one order of magnitude lower than the fitted ZFS value and the same applies to the calculated orbital magnetic moments $m_L$, as seen in Table 2. We note in passing that DFT calculations typically underestimate the values of the orbital moments when electron correlation effects are important, as it is the case of the Co atoms in the CoQDI chains. Indeed, the $m_L$ values are anisotropic and follow the same trend as the MAE as expected.[34,35] In the case of CrQDI, however, the calculated $m_L$ value is close to the one from the multiplet model.

**Scanning Probe Microscopy**

In order to provide a more detailed insight into the magnetic properties of the TM-QDI chains on atomic scale, we have also carried out low-temperature UHV scanning-probe microscopy with functionalized tips to locally probe their magnetic structure.

Figure 5 shows simultaneous atomic-force and scanning-tunneling microscopy (AFM/STM) images of CoQDI and CrQDI chains on the Au(111) surface, acquired with a CO-terminated tip. The high-resolution AFM images disclose very similar chemical composition of CoQDI and CrQDI chains, which consist of four-fold coordinated transition-metal atoms (Co, Cr) connected to the ligands via coordination bonds with four nitrogen atoms of QDI, see Figure 1b. We have carried out differential conductance dI/dV(V) point spectroscopy measurements, inspecting a possible low-bias magnetic signal with metallic tip. While the dI/dV point spectra recorded on top of a Co atom in CoQDI are featureless in a range of 60 mV (see Figure 5c), we find a zero-bias peak in the *dI/dV* spectra taken on top of a Cr atom in CrQDI (Figure 5f). The presence of the zero-bias peak indicates an underscreened Kondo effect,[38,39] having in mind the HS state of Cr ($S = 2$) previously confirmed by the XMCD analysis and DFT calculations. We find a relatively large Kondo temperature $T_K \sim 35$ K, obtained by fitting the Frota function (Figure S12),[40] which is unusually high for the underscreened Kondo effect in the HS $S = 2$ state. We tentatively assign the origin of the Kondo effect, in this particular case, to the combination of a very small orbital moment (and, thus, low magnetic anisotropy) with the low $D_{2h}$ symmetry of the Cr atom in the CrQDI wire. The net outcome is a strong mixing of $m_S = \pm 2$ and $m_S = \pm 1$ Cr states via quantum tunneling (see discussion later and Figure 8). A more elaborated theoretical analysis employing many-body techniques would be required to provide a more detailed understanding of the origin of this underscreened Kondo effect in CrQDI chains, but this task is beyond the scope of this paper.

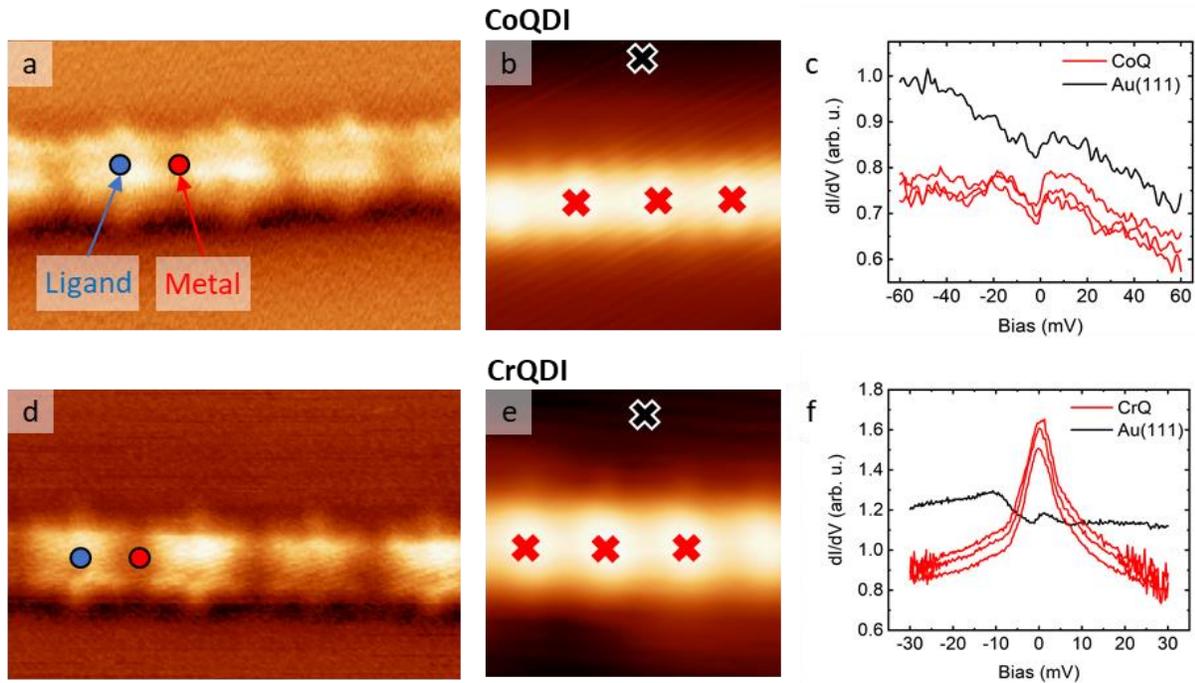

**Figure 5. AFM/STM images and dI/dV spectroscopy of CoQDI and CrQDI chains.** High-resolution constant-height AFM images of (a) CoQDI and (b) CrQDI with a CO terminated scanning-probe tip. Metal and ligand positions are highlighted. *dI/dV* measurements performed on metal sites of (c) CoQDI and (d) CrQDI with metal-terminated scanning probe including corresponding reference spectra acquired on the Au(111) surface. The corresponding point-spectroscopy positions are highlighted with red crosses in (b, e). The scanning parameters were: (a, b) Oscillation amplitude: 50 pm, bias voltage: 5 mV. Size: (a, b) $3 \times 1$ nm$^2$.

To further explore the local magnetic order, we have performed dI/dV point spectroscopy of the CoQDI and CrQDI chains using nickelocene (NiCp$_2$) functionalized tips. Unlike the square-planar 3d$^8$ Ni complexes, which are usually in a low-spin state,[41] the NiCp$_2$ molecule has the Ni atom in a high-spin ($S = 1$) triplet state ($m_S = -1, 0, +1$, *cf.* Figure 1c),[42] with an axial crystal field. The ground state is $m_S = 0$, i.e., NiCp$_2$ has an easy plane of magnetization with positive magnetic anisotropy parameter $D \sim 4$ meV.[43,44] After deposition on the SPM tip, NiCp$_2$ retains its spin $S = 1$ and magnetic anisotropy in a reproducible way.[43] The excitation from the in-plane spin $m_S = 0$ ground state to the doubly degenerate spin-up and spin-down ($m_S = \pm 1$) excited states appears in the IETS dI/dV spectra as a step at ~4 mV of voltage bias. The presence of NiCp$_2$ on the tip also substantially reduces the chemical reactivity of the metallic tip apex, something that facilitates stable tunneling conditions, even at very close tip-sample distances.[24]

We acquired a series of $d^2I/dV^2$ spectra at different tip-sample distances in the bias range of ±10 mV, which enabled us to analyze the evolution of the magnetic signal of the NiCp$_2$ tip upon site-dependent interaction with the TM-QDI chain. These height-dependent data sets also permit a direct comparison with the theoretical IETS modelling.

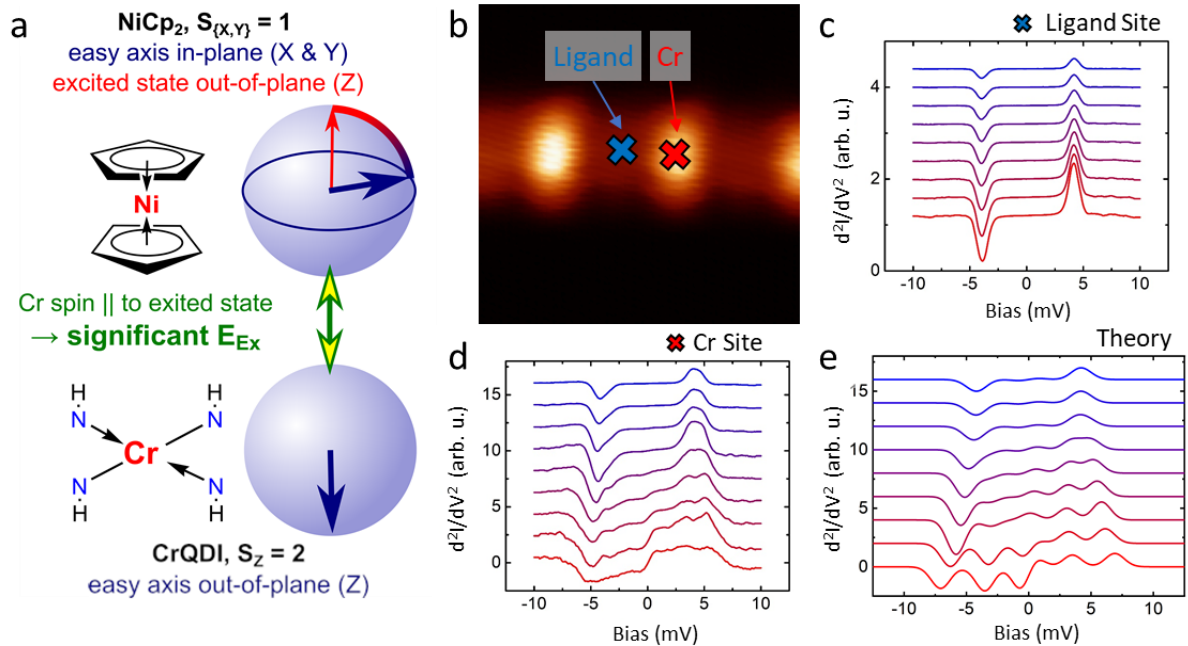

**Figure 6 Experimental and theoretical $d^2I/dV^2$ NiCp$_2$ IETS of CrQDI.** (a) Sketch of the NiCp$_2$ IETS experiment. Because of its out-of-plane easy axis, the spin of Cr is collinear to the excited state of NiCp$_2$. Hence, the ground state of Cr and the excited state of NiCp$_2$ are significantly exchange coupled, experiencing an exchange interaction energy ($E_{Ex}$). The combination of this exchange interaction with the small magnetic anisotropy energy of the Cr spin leads to a rich IETS spectra. (b) Constant-height STM image of CrQDI. (c, d) Series of $d^2I/dV^2$ spectra taken with a NiCp$_2$ terminated scanning-probe tip on the Cr and ligand sites at different tip-sample distances (10 pm between spectra). The red curves correspond to the closest approach with the STM set point of 7.7 nA at 10 mV on the Cr site. The spectroscopy position is indicated by red (Cr site) and blue (ligand site) crosses in (b). (e) Theoretical simulation of the $d^2I/dV^2$ spectra taken on Cr site corresponding to the experiment.

IETS spectra recorded with a NiCp$_2$ decorated tip on the metal (Cr) and ligand site of a CrQDI wire are shown in Figure 6c,d, respectively. The $d^2I/dV^2$ spectra of the ligand site remain almost unaltered but the $d^2I/dV^2$ spectra recorded on the Cr site show substantial changes upon the tip approach. The individual $d^2I/dV^2$ peaks located at ~4 mV, observed at long distances, first split and then progressively broaden at even shorter tip-sample distances. This strong spatial variation of the $d^2I/dV^2$ spectra on Cr-QDI chain gives evidence that the spin of NiCp$_2$ interacts with the spin of CrQDI. The $S = 2$ quintet state of Cr in CrQDI, with its out-of-plane easy axis of magnetization and small magnetic anisotropy energy, results in rather complex IETS spectra (see the detailed discussion below).

A quite different scenario is found in the case of CoQDI (Figure 7). Above both the ligand and Co site, we see IETS spectra of NiCp$_2$ tip very similar to those recorded on the metallic substrate (Figure S13) and none of them are altered upon the tip approach towards the chain (see Figure 7c,d). This indicates weak interaction of the magnetic ($S = 1$) state of the NiCp$_2$ tip with the local in-plane spin of CoQDI.

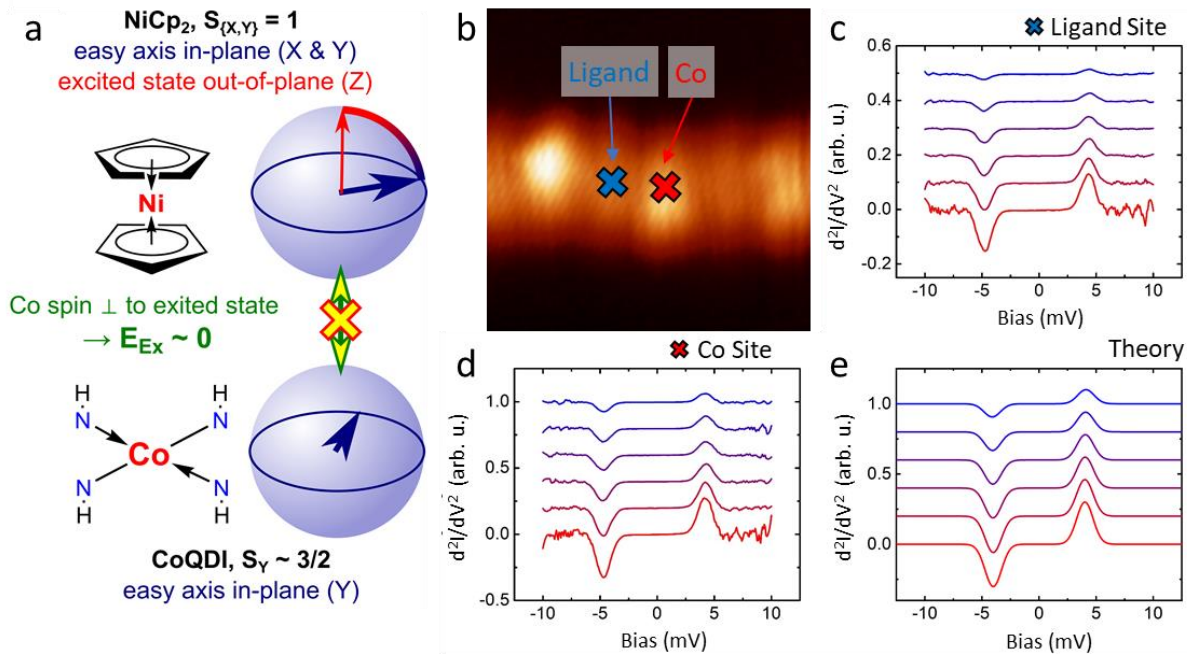

**Figure 7 Experimental and theoretical $d^2I/dV^2$ NiCp$_2$ IETS of CoQDI.** (a) Sketch of the NiCp$_2$ IETS experiment. In this simplified picture, the strong magnetic anisotropy in an in-plane direction keeps the spin of Co perpendicular to the excited state of NiCp$_2$ (which is out-of-plane). Hence, NiCp$_2$ experiences a vanishing exchange interaction energy ($E_{Ex}$) and the energy threshold of the spin excitation ($m_S = 0 \rightarrow m_S = \pm 1$) remains unchanged. (b) Constant-height STM image of CoQDI. (c, d) Series of $d^2I/dV^2$ spectra taken with NiCp$_2$ terminated scanning-probe tip on Co and ligand sites at different tip-sample distances (20 pm between spectra). The red curves correspond to the closest approach with STM a set point of 3.62 nA at 10 mV on the Co site. The spectroscopy

position is designated by red (Co site) and blue (ligand site) crosses in (b). (e) Theoretical simulation of the $d^2I/dV^2$ spectra taken on Co site corresponding to the experiment.

Based on the X-ray spectroscopy and the DFT results, the character of $d^2I/dV^2$ spectra can be understood both qualitatively and quantitatively. We recall that Co in CoQDI is characterized by a spin $S \sim 3/2$ state with a strong easy axis of magnetization along the Y direction (*i.e.* an in-plane direction) while Cr in CrQDI has spin $S = 2$ and an easy-axis of magnetization along Z (*i.e.* out of plane). Qualitatively, NiCp$_2$ IETS spectra are affected by the exchange interaction of the probed nanoobject (here CoQDI and CrQDI) with the excited $m_S = \pm 1$ states of NiCp$_2$ (*cf.* simplified sketches in Figure 6a and 7a). In the case of CrQDI, the collinear spin-alignment of the out-of-plane excited states of NiCp$_2$ with the out-of-plane ground-state of CrQDI allows an efficient magnetic exchange interaction and, hence, the NiCp$_2$ IETS spectra are modified when the tip approaches the Cr atom.

In the case of CoQDI, we observe a negligible variation of IETS spectra for two reasons. First, the magnetic anisotropy of CoQDI (ZFS $\sim$ 37 meV) is much higher compared to NiCp$_2$ (ZFS $\approx$ 4 meV) and the Co spin is effectively locked to point along the Y direction. Second, this in-plane spin state of CoQDI is orthogonal to the out-of-plane excited state of NiCp$_2$. This makes the exchange interaction between NiCp$_2$ and CoQDI negligible, *i.e.* the NiCp$_2$ IETS spectra remain almost unchanged when the tip approaches the Co atom.

Unlike the X-ray spectroscopy measurements, which probe the net magnetic moment of the Cr or Co atoms in the chain, NiCp$_2$ IETS measures the local interaction between a single Co or Cr atom and the NiCp$_2$ molecule at the tip. Hence, the presence of intra-wire exchange interactions is not expected to directly affect the IETS signal. In the case of the in-plane easy axis of magnetization (Co), strong FM or AFM intra-wire exchange interactions could, however, increase the energy required for spin canting. This would further reduce the projection of the spin on the Z axis and hence the splitting of the NiCp$_2$ m$_S$ = ±1 states.

**IETS Simulations**

To gain a quantitative understanding, we have developed an IETS model formulated in terms of the electronic degrees of freedom, which allows for a sufficiently detailed description of the electron tunneling through the magnetic system (the NiCp$_2$ molecule coupled to the CoQDI or CrQDI chain) located between the metallic electrodes (tip and substrate). We have assumed that the magnetic system can be reduced to a two-site Hubbard model consisting of the partially filled 3d shell in the NiCp$_2$ molecule and another 3d shell in the probed transition-metal atom in the TM-QDI chain. The Hubbard Hamiltonian includes the crystal field, the spin-orbit coupling and the Coulomb interaction in each 3d shell, as well as an effective hopping *t* between them. Increasing hopping *t* corresponds to the NiCp$_2$-tip approaching the TM-QDI chain. The crystal field and spin-orbit coupling are taken from the multiplet model fitted to the x-ray absorption spectra. The tunneling current is calculated using the cotunneling theory.[45] A detailed description of the IETS model is provided in the Methods section.

In the case of CoQDI, the simulated IETS spectra in the ±10 mV range remain unchanged despite increasing the hopping parameter $t$ (Figure 7e), which is consistent with the experimental data. The IETS spectra remain almost unaltered due to the perpendicular alignment of the spin of the excited state of NiCp2 tip and the spin ground state of the Co site. Only at very close distances, there is a slight upward movement of the first excited state of NiCp2 tip to higher energies as can be seen in Figure S14 that shows the evolution of the low-energy excitations of the Hubbard model as a function of $t$.

Figure 6e displays the simulated IETS spectra for CrQDI, showing a significant broadening and subsequent splitting of the IETS peaks as the tip approaches the CrQDI chain, reproducing rather well the experimental IETS spectra.

It is instructive to analyze the IETS spectra using the (spin) Heisenberg model in the form

$$\widehat{H} = D_{\mathrm{Ni}}\left(\widehat{S}_z^{\mathrm{Ni}}\right)^2 + D_{\mathrm{Cr}}\left(\widehat{S}_z^{\mathrm{Cr}}\right)^2 + J\widehat{\mathbf{S}}^{\mathrm{Ni}} \cdot \widehat{\mathbf{S}}^{\mathrm{Cr}}$$

represents the low-energy limit of our two-site Hubbard model. Indeed, the Heisenberg model (with parameters $D_{\mathrm{Ni}} \approx 4$ meV and $D_{\mathrm{Cr}} \approx -0.23$ meV and $J \approx 0.0975 t^2$) approximates the solution of the electronic model very well, as shown on Figure 8a. The Heisenberg model has nine eigenstates $\Psi_i$, the basic characteristics of which are listed in Figure 8c. Most of the eigenstates are doubly degenerate, three of them are singlets. All quadratic (and higher-order) terms in $J$ disappear from the eigenenergies when the Heisenberg exchange is replaced with the Ising uniaxial exchange $J\widehat{S}_z^{\mathrm{Ni}}\widehat{S}_z^{\mathrm{Cr}}$, but such a simplified model does not represent the Hubbard model very accurately. Figure 8b displays the simulated IETS spectrum of CrQDI for a hopping parameter value $t = 0.06$ eV with the assigned origin of the individual peaks. The state $\Psi_0$ represents the ground state, the peak at the lowest bias voltage (~1 mV) corresponds to $\Psi_1$ (spin flip at the Cr site), the second peak originates from $\Psi_3$ (spin flip at the Ni site), and the accidental doublet found at ~5 mV originates from $\Psi_4$ (spin flip at the Ni site) and $\Psi_7$ (spin flip at both sites).

Our Heisenberg model is axially symmetric, hence the expectation value of $\hat{S}_z^{Ni} + \hat{S}_z^{Cr}$, $m_s^{tot} = m_s^{Ni} + m_s^{Cr}$, is a good quantum number. The individual expectation values, $m_s^{Ni}$ and $m_s^{Cr}$, are not conserved, since all states with a given $m_s^{tot}$ are mixed by the exchange term. The values of $m_s^{Ni}$ and $m_s^{Cr}$ shown in Figure 8c correspond to a vanishing exchange term ($J = 0$). According to cotunneling theory, the IETS transitions starting from the ground state $\Psi_0$ are allowed to states $\Psi_1$, $\Psi_3$, $\Psi_4$ and $\Psi_7$. This corresponds to one exact selection rule $|\Delta m_s^{tot}| \leq 1$ (the electron tunneling from one electrode to the other can change its spin at most by ±1, *i.e.*, flip its spin) and to two approximate selection rules $|\Delta m_s^{Ni}| \leq 1$ and $|\Delta m_s^{Cr}| \leq 1$ (these would be exact if the spin flips at Ni and Cr sites were independent sequential events). The transition $\Psi_0 \to \Psi_8$ is forbidden only by one of the approximate rules, $|\Delta m_s^{Cr}| \leq 1$. Indeed, a closer inspection of the numerical data reveals a small contribution.

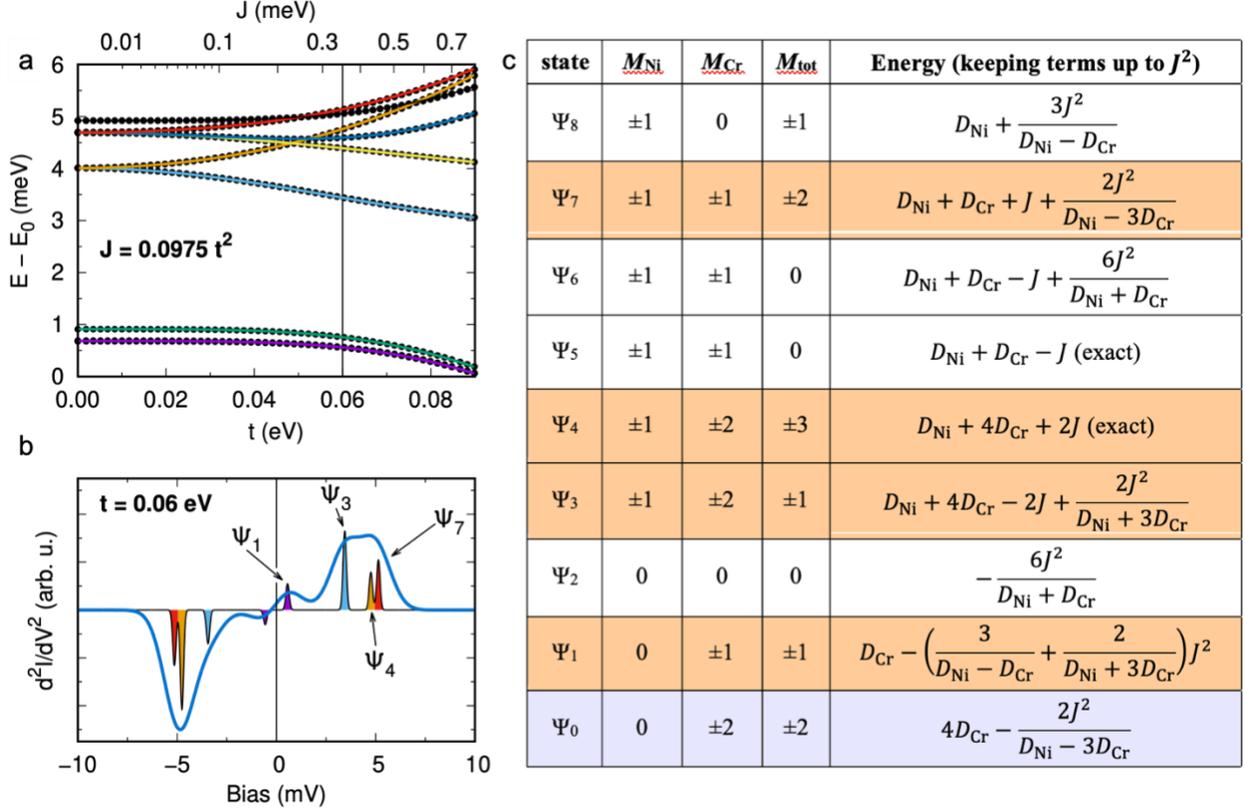

**Figure 8 Theoretical analysis of $d^2I/dV^2$ spectrum of NiCp$_2$ tip and CrQDI with assigned origin of peaks.** a) Evolution of the excitation energies with respect to the ground-state energy $E_0$ calculated by Hubbard (dots) and Heisenberg (lines) models. b) Calculated $d^2I/dV^2$ spectrum of NiCp$_2$ tip and CrQDI with assigned origin of the individual peaks. c) Eigenstates and eigenenergies (approximate values up to 2$^{nd}$ order in the exchange coupling $J$) of the Heisenberg model representing the coupled magnetic system comprised of NiCp$_2$ tip and CrQDI. The eigenstates are ordered from bottom to top by increasing energy evaluated at $J = 0$.

**Conclusions**

Our thorough study of the magnetic properties of 1D metal-organic CrQDI and CoQDI chains on the Au(111) surface has enabled us to determine the spin magnetic moments of the Cr ($S = 2$, easy axis out of plane) and Co ($S = 3/2$, easy axis in an in-plane direction) atoms, as well as their preferred magnetization directions.

We show that IETS spectroscopy acquired with NiCp$_2$ decorated tip is a formidable characterization tool. This technique combined with the appropriate theoretical analysis can provide detailed information about the local magnetic properties. In combination with XAS/XMCD and multiplet calculations, it is possible to rationalize the absence of variations of the IETS signal of the NiCp$_2$ tip probing CoQDI chain. We hope that the detection of the underscreened Kondo effect in high-spin ($S = 2$) CrQDI chain with an unusually high Kondo temperature will stimulate future in-depth theoretical investigation to explain its origin.

## Methods

### Sample Preparation

For all experiments, the samples were prepared in-situ in UHV conditions. The single-crystalline Au(111) substrates were prepared by repeated cycles of Ar$^+$ etching and annealing. The precursor molecule 2,5-diamino-1,4-benzoquinonediimine (2HQDI) was synthesized using the procedure described in the literature.[3] The molecules were sublimed from a tantalum pocket heated to 115°C. The chains were grown by co-deposition of precursor molecule with the respective metal atom (Co or Cr) onto a substrate kept at elevated temperature (300°C).[26]

### SPM Experiments

The SPM experiments were carried out in a commercial low-temperature (1.2 K) STM/nc-AFM microscope (Specs-JT Kolibri sensor: $f \sim 1$ MHz). Nickelocene (NiCp$_2$) was deposited from a tantalum pocket kept at room temperature onto the prepared sample in the microscope head kept at < 4 K. Prior to the tip functionalization, the symmetricity of the metallic tip apex was checked by imaging the characteristic round shape of the Nc molecule. To pick the NiCp2, a sample area containing multiple molecules was scanned at low current and bias set point (< 50 pA, 1 mV) until spontaneous functionalization occurred. Before nickelocene IETS, we confirmed that the *dI/dV* spectra of the metallic tip are featureless. After the functionalization of the metallic tip with NiCp2, *d²I/dV²* spectra recorded on the bare Au(111) reveal two IETS peaks at ~3.9 meV. The stability of functionalized NiCp2 tip was tested by *dF/z* spectroscopy – as stable were considered tips exhibiting no hysteretic behavior for forward and backward curves. All the *dI/dV* spectra were taken with an external lock-in amplifier at 1.2 K with 0.5 mV (rms) modulation voltage.

**X-ray Absorption Measurements**

The X-ray absorption spectroscopy (XAS) measurements were performed at the X-Treme beamline[46] of the Swiss Light source at 3 K in total-electron yield mode using circularly ($\sigma^+$ and $\sigma^-$) and linearly ($\sigma^h$ and $\sigma^v$) polarized X-rays. XMCD and XLD correspond to the differences ($\sigma^+ - \sigma^-$) and ($\sigma^h - \sigma^v$), respectively, while XAS corresponds to the sums ($\sigma^+ + \sigma^-$) and ($\sigma^h + \sigma^v$). The XMCD spectra were recorded in normal and grazing (60° off the surface normal) incidence at an applied magnetic field of 6.8 T. The XLD spectra were recorded at an applied field of 0.1 T in grazing incidence. The applied magnetic field was parallel to the X-ray incidence direction. The spectra were normalized to unity at the pre-edge and the XAS background recorded on clean Au(111) was subtracted.

**Multiplet Calculations**

Multiplet calculations were performed using the multiX code[27] which treats the crystal field as electrostatic potential of effective point charges. Here, the point charges placed at the positions of the nitrogen and carbon atoms as obtained by DFT are considered as free parameters. Thus, the $D_{2h}$ symmetry of the MQDI wires is respected. Additional free parameters are the scaling factors of the spin-orbit and Coulomb interactions that account for the screening of the free-ion quantities. The free parameters are optimized (values in Tables S1 and S2) by minimizing the least squares error between the simulated and measured grazing incidence XAS and XLD spectra. To reproduce the in-plane disorder caused by the wires being aligned along different directions on the Au(111) surface,[26] the simulated spectra are calculated by summing spectra obtained for different in-plane angles (6 per 180° sector).

**DFT calculations**

Density functional theory (DFT) calculations that involved the GGA+U functional were carried out for free-standing CoQDI and CrQDI chains using the supercell approach, with the one-dimensional (1D) chains lying along the X-axis with 12 Å lateral distance (Y-axis) separation, and the Z-axis being perpendicular to the plane containing the CoQDI or CrQDI metal-organic units. The VASP code, based on plane-wave basis sets and projector augmented potentials,[47–49] was used with a 450 eV cut-off for the plane waves and a $10 \times 1 \times 1$ sampling of the supercell Brillouin zone.[50] The structures were relaxed with a 0.01 eV/Å tolerance in forces and $10^{-7}$ eV in the total energies. The calculations were done with the PBE exchange-correlation functional[51,52] augmented with DFT+U[28,29] corrections to model electron correlation effects in the partially filled 3d shells using Dudarev approximation $U_{eff} = U - J$. In order to analyze the dependence of the electronic structure on the $U$ parameter and to search for the global minimum among the multiple local minima of the DFT+U total energy, the occupation matrix control (OMC) method was used.[53]

A vibrational analysis of the chains was carried out in the harmonic approximation, by diagonalization of the dynamical matrix (calculated by finite differences) at the Gamma point.

Exchange coupling constants $J$ were calculated by doubling the unit cell and imposing ferro (FM) and antiferromagnetic (AFM) order of the spins. Additionally, the distance dependence of $J$ was obtained from spin spiral calculations, modelled in the generalized Bloch theorem approach for $q$ wavevectors defining the spiral wave length ($2\pi/q$), by a Fourier transform of the so-obtained $J(q)$ values.[54,55] To obtain magneto crystalline anisotropy energies (MAE) and orbital magnetic moments, the spin-orbit coupling contribution was treated self-consistently and also perturbatively within the force theorem (FT) approach.[56,57]

Hybrid DFT calculations were performed with FHI-AIMS code[58] using PBE0[30] exchange-correlation functional with the contribution of the exact exchange component varied from 0.25 (default value) to 0.4. In all the calculations, we employed the numerical atomic orbital basis sets. The atomic structures were relaxed until the total forces reached below $5\times10^{-3}$ eV Å$^{-1}$. The Brillouin zone was sampled with 12 $k$-points along the periodic axis of the TM-QDI chains for a doubled unit cell (needed for the antiferromagnetic calculation).

**IETS Model**

For the purpose of modeling the inelastic electron tunneling, we assume that all relevant processes occur in the magnetic centers, namely in their partially filled 3d shells. Hence, we work in a reduced model consisting of these 3d shells, each described by a Hamiltonian $\hat{H}_i$, and of effective couplings $\hat{T}_{ij}$ between them

$$\hat{H}_i = \sum_{\sigma\sigma'mm'} \left( \zeta_i [l \cdot s]_{mm'}^{\sigma\sigma'} + h_{imm'}\delta_{\sigma\sigma'} + \epsilon_i \delta_{mm'}\delta_{\sigma\sigma'} \right) \hat{d}_{im\sigma}^\dagger \hat{d}_{im'\sigma'} + \hat{U}_i,$$

consists of the spin-orbit coupling $\zeta_i$, the crystal-field potential $h_i$ and the Coulomb interaction $\hat{U}_i$ that is parametrized by three Slater integrals $F_0^i$, $F_2^i$ and $F_4^i$. The matrix $h_i$ is assumed to be traceless and the energy of the 3d level is introduced separately as $\epsilon_i$. Index $i$ distinguishes individual 3d shells, $\sigma$ is the spin index, $m$ is the magnetic quantum number, and $\hat{d}^\dagger_{im\sigma}$ is the creation operator that inserts an electron into the spinorbital with these indices. The single-shell Hamiltonian $\hat{H}_i$ has exactly the same structure as the Hamiltonian employed in the multiX code to simulate the X-ray absorption spectroscopies, more precisely, their initial state. Thus, we can directly use the numerical values of $h_i$ and $\zeta_i$ obtained by fitting the X-ray spectra also for the IETS modeling (Table S7). The parameters $F_k^i$ are slightly different due to differences in software implementations.

The two 3d shells are coupled by a hopping operator

$$\hat{T}_{12} = \sum_{\sigma m}(t_m \hat{d}^\dagger_{1m\sigma}\hat{d}_{2m\sigma} + t_m^* \hat{d}^\dagger_{2m\sigma}\hat{d}_{1m\sigma}),$$

which we assume to be axially symmetric around the $z$ axis so that the magnetic quantum number $m$ (as well as spin $\sigma$) is conserved when an electron hops from one shell to the other. For simplicity, we neglect the orbital dependence of the hopping amplitudes and choose $t_m = t$. A more realistic setting could be derived from (or at least inspired by) the DFT electronic structure

In the tunneling experiment, some of the $\hat{d}_{im\sigma}$ states are coupled to the electrodes, one of which is the STM tip and the other is the substrate. In our setup, the Ni 3d shell in the nickelocene is coupled to appropriate linear combinations[59,60] of the conduction-electron states $\hat{a}_{km\sigma}$ in the tip. The coupling is described by a Hamiltonian term

$$\hat{V}_{tip} = \sum_{km\sigma}(v_{km}\hat{d}^\dagger_{1m\sigma}\hat{a}_{km\sigma} + v_{km}^*\hat{a}^\dagger_{km\sigma}\hat{d}_{1m\sigma}).$$

An analogous term is added for coupling of the other 3d shell to the conduction states $\hat{b}_{km\sigma}$ in the substrate,

$$\hat{V}_{sub} = \sum_{km\sigma}(w_{km}\hat{d}^\dagger_{2m\sigma}\hat{b}_{km\sigma} + w^*_{km}\hat{b}^\dagger_{km\sigma}\hat{d}_{2m\sigma}).$$

As written, the operators $\hat{V}_{tip}$ and $\hat{V}_{sub}$ are axially symmetric around the $z$ axis like the operator $\hat{T}_{12}$. Similarly to the inter-shell hopping amplitudes $t_m$, we simplify the tip and substrate hopping amplitudes to $v_{km} = v$ and $w_{km} = w$ in the actual calculations.

The tunneling current is computed using cotunneling theory[45], which corresponds to the lowest-order of the perturbation expansion in the coupling to the tip and substrate, $\hat{V}_{tip}$ and $\hat{V}_{sub}$. Keeping only the lowest-order contributions means that the tunneling through the magnetic system between the tip and substrate proceeds one electron (or one hole) at a time. It also means that Kondo effect is neglected, since it is a higher-order effect.

After a tunneling event, the magnetic system can stay in an excited state, which is the essence of the inelastic tunneling. In addition to the approximations described in ref. 45, we assume that the system has enough time to relax back to the ground state before the next tunneling event starts. Finally, we neglect all thermal effects by setting the temperature to 0 K.

The electronic structure of the electrodes enters the formula for the differential conductance only through their densities of the electronic states (DOS). Given that we explore only very small biases (< 10 mV), which corresponds to probing the densities of states only ±10 meV around the Fermi level, the DOSes can be replaced by constants. Hence, the characteristics of the electrodes do not affect the shape of the IETS spectra.

ASSOCIATED CONTENT

**Supporting Information Available:** Parameters for the multiplet calculations and M(H) data, DFT+U calculations, hybrid DFT calculations at the PBE0 level with different amount of Fock exchange, additional STM spectroscopy data and fit of the Kondo resonance, and parameters of simulations of IETS in the Hubbard model (PDF).

AUTHOR INFORMATION

**Corresponding Authors**

* christian.waeckerlin@empa.ch, kolorenc@fzu.cz, andres.arnau@ehu.eus, jelinekp@fzu.cz

**Author Contributions**

The manuscript was written through contributions of all authors. All authors have given approval to the final version of the manuscript.

**Funding Sources**

Swiss National Science Foundation (Grants P300P2_177755 and 173720)

University of Zürich Research Priority Program LightChEC

Spanish PID2019-103910GB-I00, funded by MCIN/AEI/10.13039/501100011033/ and FEDER *Una manera de hacer Europa*, as well as GIU18/138 by Universidad del País Vasco UPV/EHU; IT-1246-19and IT-1260-19 by Gobierno Vasco

Mobility Plus Project No. CNR-19-03 of the Czech Academy of Sciences

Praemium Academie of the Academy of Science of the Czech Republic

Czech Science Fundation projects no. 20-13692X and 20-18741S

Czech Nanolab Research Infrastructure supported by MEYS CR project no. LM2018110

ACKNOWLEDGMENT

C.W. gratefully acknowledges financial support by the Swiss National Science Foundation (Grants P300P2_177755 and 173720) and the University of Zürich Research Priority Program

LightChEC. We thank J. Dreiser for support and helpful discussions. J.G., M.B-R and A.A. acknowledge DIPC supercomputing center. J.K. acknowledges financial support by the Czech Academy of Sciences (Mobility Plus Project No. CNR-19-03). We acknowledge support of the Praemium Academie of the Academy of Science of the Czech Republic and the Czech Nanolab Research Infrastructure supported by MEYS CR (project no. LM2018110). P.J., M.O., A.C. acknowledge support of GACR project no. 20-13692X. M.S. acknowledges support of GACR project no. 20-18741S.

# Supporting Information

# The Role of the Magnetic Anisotropy in Atomic-Spin Sensing of 1D Molecular Chains

*Christian Wäckerlin,[a,b] Aleš Cahlík,[a] Joseba Goikoetxea,[c] Oleksandr Stesovych,[a] Daria Medvedeva,[d] Jesús Redondo,[a] Martin Švec,[a] Bernard Delley,[e] Martin Ondráček,[a] Andres Pinar,[a] Maria Blanco-Rey,[f,g] Jindrich Kolorenc,[d] Andres Arnau,[c,f,g] Pavel Jelínek[a,g]*

[a] Institute of Physics, Czech Academy of Sciences, Cukrovarnická 10, 16200 Prague, Czech Republic

[b] Surface Science and Coating Technologies, Empa, Swiss Federal Laboratories for Materials Science and Technology, Überlandstrasse 129, 8600 Dübendorf, Switzerland

[c] Centro de Física de Materiales CFM/MPC (CSIC-UPV/EHU), Paseo Manuel de Lardizábal 5, 20018 Donostia-San Sebastián, Spain

[d] Institute of Physics, Czech Academy of Sciences, Na Slovance 2, 18221 Prague, Czech Republic

[e] Condensed Matter Theory, Paul Scherrer Institut, CH-5232 Villigen, Switzerland

[f] Departamento de Polímeros y Materiales Avanzados: Física, Química y Tecnología, Facultad de Química UPV/EHU, Apartado 1072, 20080 Donostia-San Sebastián, Spain

[g] Donostia International Physics Center (DIPC), Paseo Manuel de Lardizábal 4, 20018 Donostia-San Sebastián, Spain

# Parameters for the multiplet calculations and M(H) data

**Table S1. Effective point charges for CrQDI.** Additional parameters are: Ground state: $3d^4$, spin-orbit scaling factor: 0.968, Coulomb scaling factor: 1.062

| X (Å) | Y (Å) | Z (Å) | Charge ($q_e$) | Atom label |
|---|---|---|---|---|
| -3.99363 | 1.43522 | 0 | -0.0006 | C |
| -5.20394 | 0.74795 | 0 | 2.4679 | C |
| -5.20394 | -0.74795 | 0 | 2.4679 | C |
| -3.99363 | -1.43522 | 0 | -0.0006 | C |
| -2.78331 | -0.74795 | 0 | 2.4679 | C |
| -2.78331 | 0.74795 | 0 | 2.4679 | C |
| -1.57349 | 1.27598 | 0 | -2.1663 | N |
| -1.57349 | -1.27598 | 0 | -2.1663 | N |
| -6.41375 | 1.27598 | 0 | -2.1663 | N |
| -6.41375 | -1.27598 | 0 | -2.1663 | N |
| 3.99366 | 1.43522 | 0 | -0.0006 | C |
| 2.78334 | 0.74795 | 0 | 2.4679 | C |
| 2.78334 | -0.74795 | 0 | 2.4679 | C |
| 3.99366 | -1.43522 | 0 | -0.0006 | C |
| 5.20397 | -0.74795 | 0 | 2.4679 | C |
| 5.20397 | 0.74795 | 0 | 2.4679 | C |
| 6.41379 | 1.27598 | 0 | -2.1663 | N |
| 6.41379 | -1.27598 | 0 | -2.1663 | N |
| 1.57352 | 1.27598 | 0 | -2.1663 | N |
| 1.57352 | -1.27598 | 0 | -2.1663 | N |

**Table S2. Effective point charges for CoQDI.** Additional parameters are: Ground state: $3d^7$, spin-orbit scaling factor: 0.881, Coulomb scaling factor: 0.931

| X (Å) | Y (Å) | Z (Å) | charge ($q_e$) | Atom label |
|---|---|---|---|---|
| -6.37583 | 1.27836 | 0 | -3.5373 | N |
| -6.37583 | -1.27836 | 0 | -3.5373 | N |
| -5.17783 | 0.75148 | 0 | -0.3325 | C |
| -5.17783 | -0.75148 | 0 | -0.3325 | C |
| -3.96755 | 1.43281 | 0 | -0.0038 | C |
| -3.96755 | -1.43281 | 0 | -0.0038 | C |
| -2.75732 | -0.75147 | 0 | -0.3325 | C |
| -2.75732 | 0.75147 | 0 | -0.3325 | C |
| -1.55928 | 1.27833 | 0 | -3.5373 | N |
| -1.55928 | -1.27833 | 0 | -3.5373 | N |
| 1.55947 | 1.27835 | 0 | -3.5373 | N |
| 1.55947 | -1.27835 | 0 | -3.5373 | N |
| 2.75748 | 0.75147 | 0 | -0.3325 | C |
| 2.75748 | -0.75147 | 0 | -0.3325 | C |
| 3.96774 | 1.4328 | 0 | -0.0038 | C |
| 3.96774 | -1.4328 | 0 | -0.0038 | C |
| 5.178 | -0.75148 | 0 | -0.3325 | C |
| 5.178 | 0.75148 | 0 | -0.3325 | C |
| 6.37601 | 1.27836 | 0 | -3.5373 | N |
| 6.37601 | -1.27836 | 0 | -3.5373 | N |

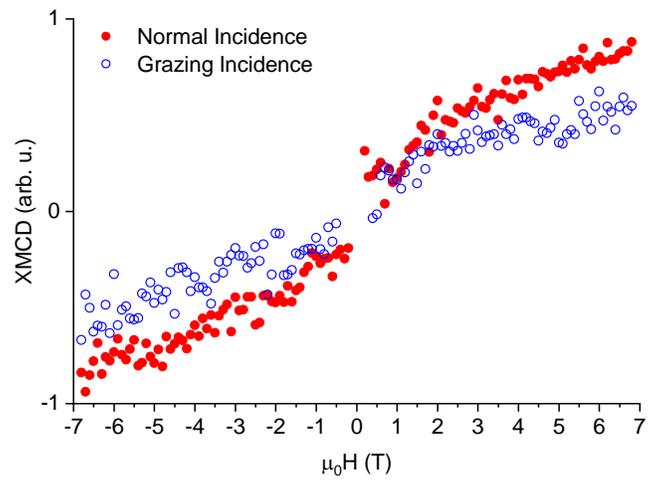

**Figure S1.** CrQDI magnetization curves recorded in grazing and normal incidence at 3 K.

## DFT+U calculations

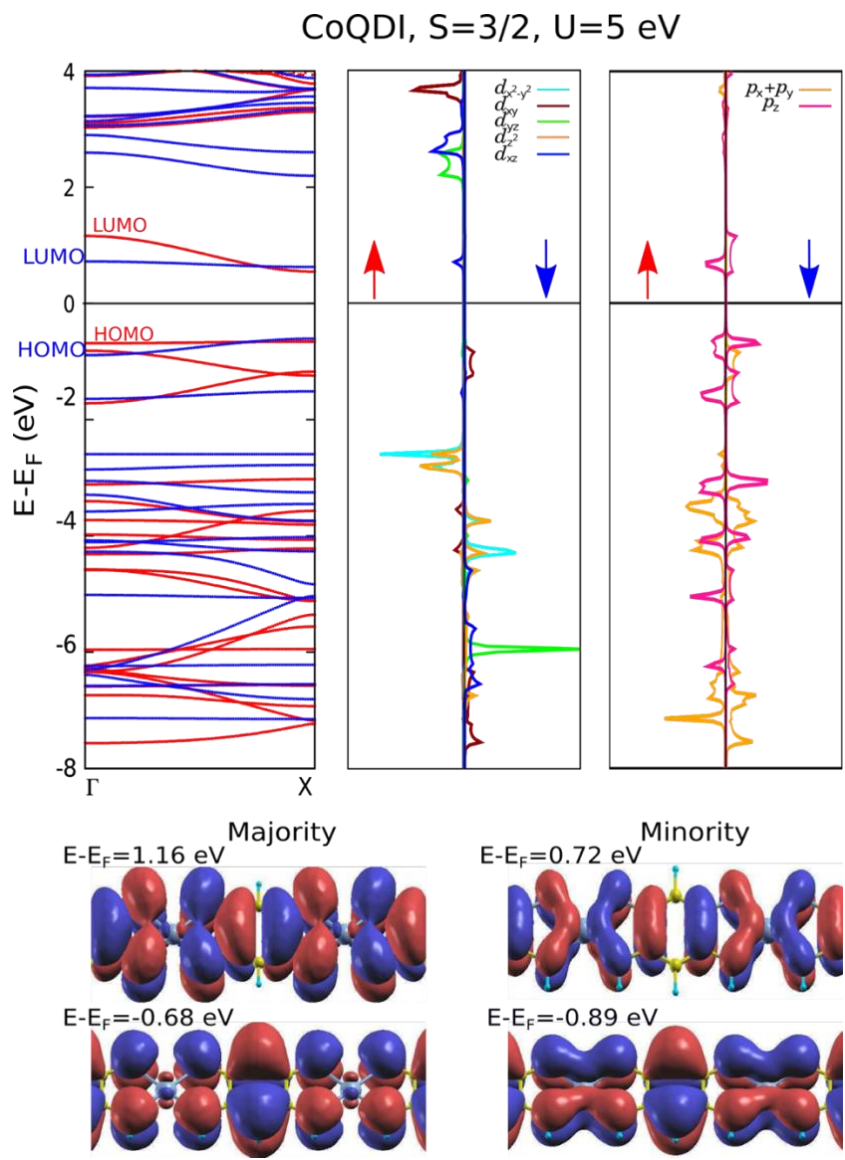

**Figure S2.** (Upper left panel) Band structure of a CoQDI 1D periodic chain for the $U = 5$ eV high spin (HS) magnetic state, where majority and minority spin bands are plotted in red and blue, respectively. (Upper middle panel) Projected density of states (PDOS) onto Co *3d* orbitals. (Upper right panel) PDOS onto N *2p* orbitals. The arrows indicate the spin majority and minority PDOS. The bottom panels show the LUMO and HOMO orbitals of both spin channels. They are obtained as the real-space representation of the Kohn-Sham states at the Γ point.

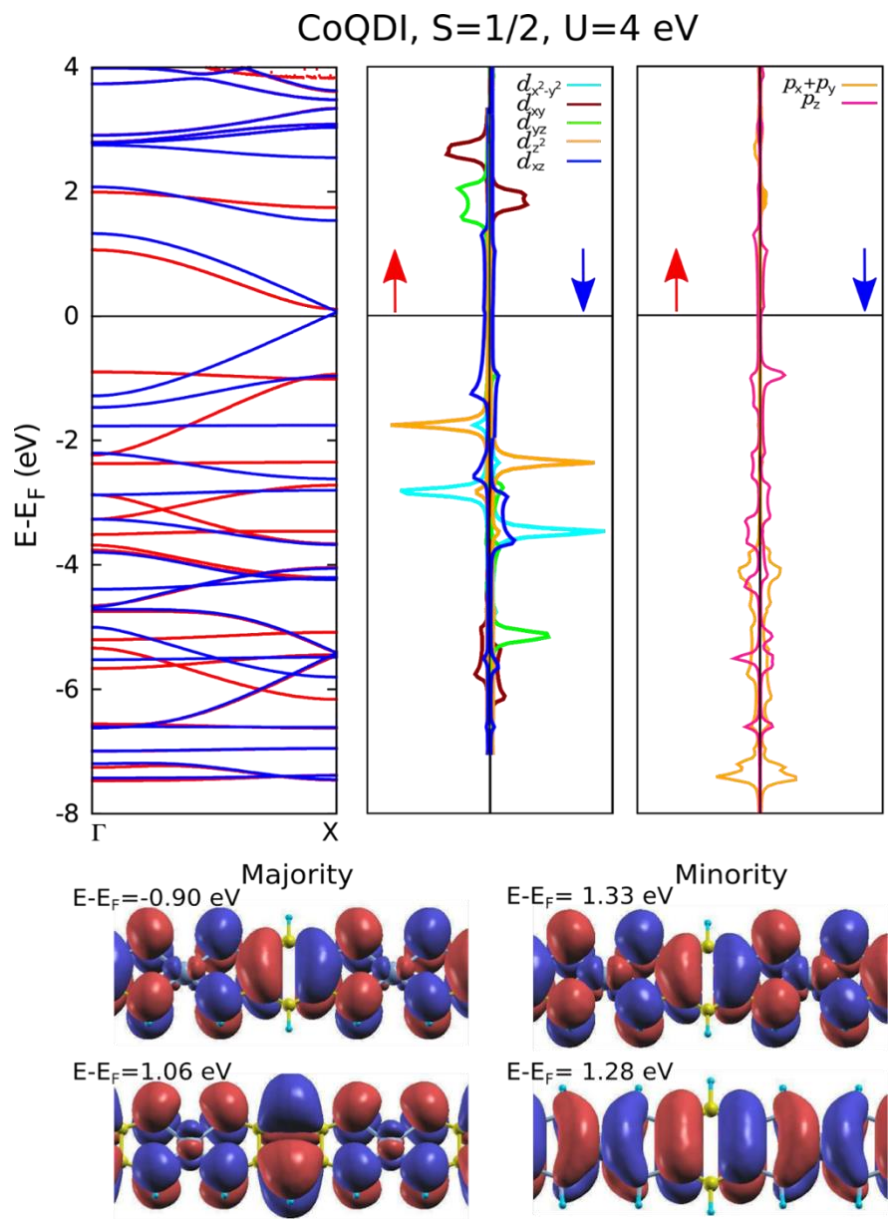

**Figure S3.** Same as Figure S2 for the $U = 4$ eV LS state of the CoQDI chain.

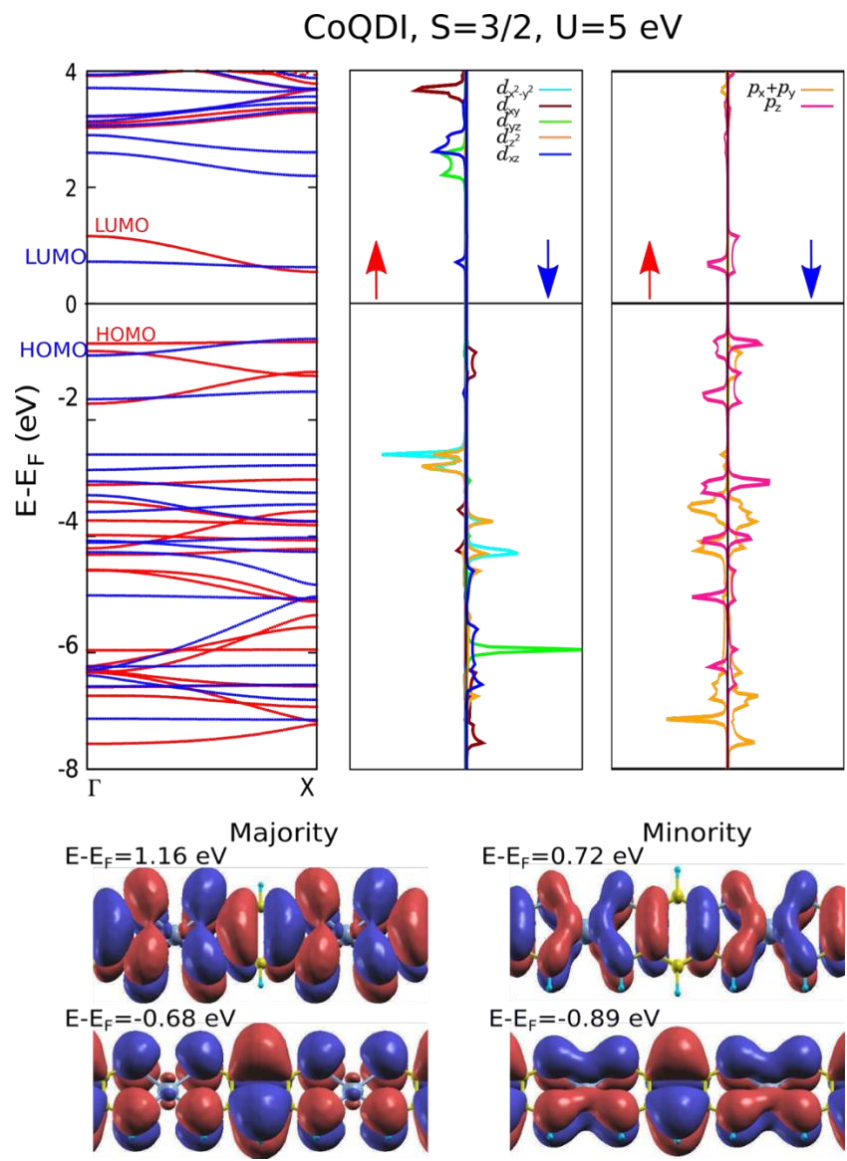

**Figure S4.** Same as Figure S2 for the CrQDI chain (U = 5 eV).

**Table S3. Exchange coupling constant (*J*) between Cr atoms in a CrQDI 1D chain for different values of the Hubbard *U* parameter.** Notice the significant decrease in J value as U increases.

| $U$ (eV) | $J$ (meV/$\mu_B^2$) |
|---|---|
| 0 | -19.03 |
| 3 | -14.56 |
| 4 | -6.45 |
| 5 | -4.42 |

**Table S4. Exchange coupling constant (*J*) between Cr atoms in a CrQDI 1D chain for *U* = 5 eV, including spin-orbit coupling (SOC), *i.e.* accounting for magnetic anisotropy, when spins are aligned along x, y or z axes as indicated.** Two different methods have been used (see Methods Section in the Manuscript): fully self-consistent (SCF) method and the perturbative approach based on the force theorem (FT). Anisotropic effects in the magnetic exchange coupling are clearly negligible (well below 1%) in this system.

| Axis | $J$ (meV/$\mu_B^2$) | |
|---|---|---|
| | SCF | FT |
| x | -4.336 | -15.160 |
| y | -4.339 | -15.161 |
| z | -4.335 | -15.161 |
| No SOC | -4.42 | |

**Table S5. CrQDI 1D chains magnetic anisotropy energy (MAE) for different $U$ values.** The MAE is defined as the total energy difference corresponding to two different magnetization directions along the *a* and *b* axis with a and b = x, y and z defined by the cartesian coordinates.

| U (eV) | MAE = $E^a$-$E^b$ (meV) | | |
|---|---|---|---|
| | x-y | x-z | y-z |
| 0 | 0.22 | 0.45 | 0.23 |
| 3 | 0.12 | 0.53 | 0.41 |
| 4 | 0.08 | 0.53 | 0.45 |
| 5 | 0.06 | 0.54 | 0.48 |

**Table S6. CrQDI 1D chains orbital moments ($m_i^{orb}$, with $i$ = x, y, z) for different $U$ values.** A direct comparison with previous data shown in Table S5 reveals that the MAE, measured as the largest difference x-z, slightly increases as U increase, and $m_z^{orb}$ follows this behavior but it is antiparallel to $m_s$, as it should be due to Hund's rule.

| U (eV) | $m_x^{orb}$ ($\mu_B$) | $m_y^{orb}$ ($\mu_B$) | $m_z^{orb}$ ($\mu_B$) |
|---|---|---|---|
| 0 | -0.018 | -0.022 | -0.028 |
| 3 | -0.011 | -0.019 | -0.033 |
| 4 | -0.012 | -0.019 | -0.036 |
| 5 | -0.013 | -0.016 | -0.038 |

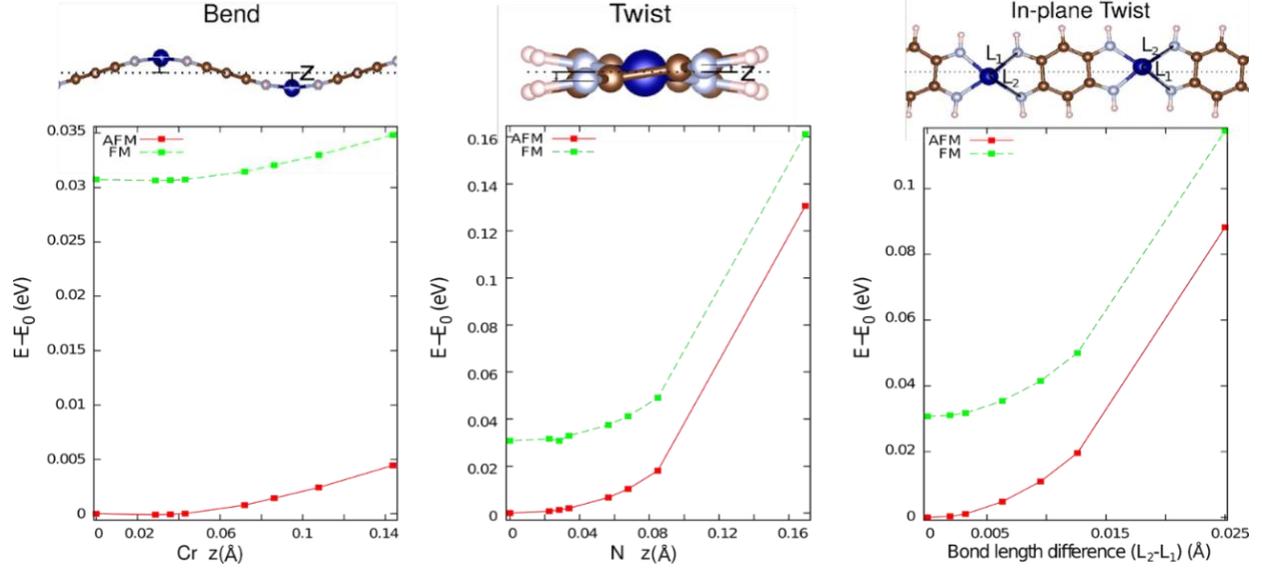

**Figure S5.** Energy difference between the AFM and FM magnetic configurations, referred to the most stable AFM configuration without any geometrical distortion, of Cr spin magnetic moments in CrQDI chains (two Cr atoms in a doubled periodic supercell). The amplitudes of the three different low-energy vibrational modes defining the geometrical distortion correspond to vertical (z coordinate) displacements for the bending and vertical twist modes, or to lateral (x and y coordinates) displacements for the in-plane twist. The corresponding eigenfrequencies (in THz) are 1.52 / [1.56], 1.27 / [1.25] and -0.73* / [0.75*] for the bending, vertical twist and in-plane twist modes, respectively, in FM / [AFM] configurations. The * symbol means imaginary value, i.e., true soft-mode. The conversion factor from THz frequency to meV energy is 1 THz = 4.13 meV. Each of them are visualized above each plot. Note that the amplitude of the distortions is enlarged in the models for a clear visualization. Interestingly, the magnetic order (AFM or FM) of the chains is preserved under the excitation of these modes.

# Hybrid DFT calculations at the PBE0 level with different amount of Fock exchange

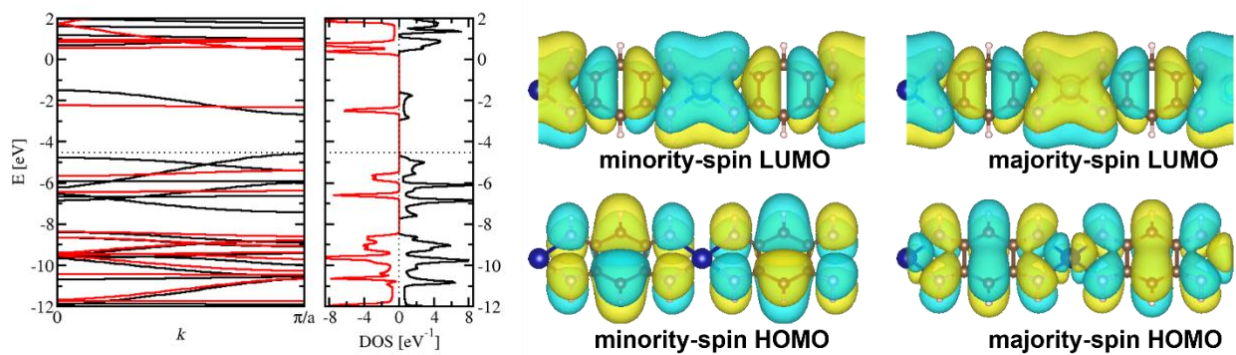

**Figure S6.** Band structure of the **CrQDI** chain (leftmost panel) in the ferromagnetic state, the corresponding density of states (next panel to the right), and frontier orbitals (more to the right). Black solid lines represent the majority spin, red solid lines minority spin states. Black dotted line denotes the Fermi level. Depicted orbitals: Minority-spin lowest-unoccupied top left; majority spin lowest-unoccupied top right; minority-spin highest-occupied bottom left; majority-spin highest-occupied bottom right. The results shown here are from the calculation with the PBE0 exchange-correlation functional.

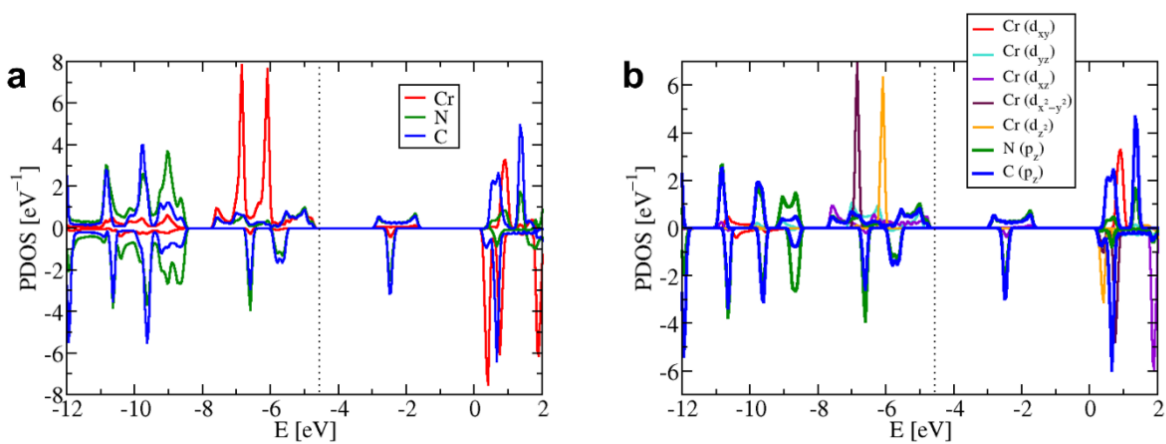

**Figure S7.** Species- and orbital-projected densities of states in the ferromagnetic **CrQDI** chain. (a) Species-projected DOS: Red lines chromium atoms, dark green lines nitrogen atoms, and blue

lines carbon atoms. (b) Orbital-projected DOS: red, turquois, violet, maroon, and orange lines stand for $d_{xy}$, $d_{yz}$, $d_{xz}$, $d_{x^2-y^2}$, and $d_{z^2-r^2/3}$ orbitals of Cr, respectively, while dark green and blue stand for $p_z$ orbitals of nitrogen and carbon atoms, respectively. Positive values represent majority spin, negative values minority spin. The results shown here are from the calculation with the PBE0 exchange-correlation functional.

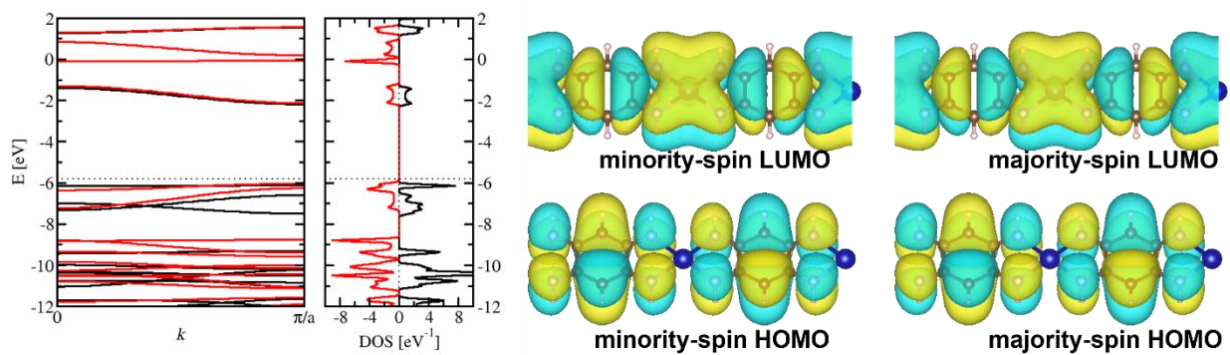

**Figure S8.** Band structure of the **CoQDI** chain in the ferromagnetic **high-spin** state (leftmost panel), the corresponding density of states (next panel to the right), and frontier orbitals (more to the right). Black solid lines represent the majority spin, red solid lines minority spin states. Black dotted line denotes the Fermi level. Depicted orbitals: Minority-spin lowest-unoccupied top left; majority spin lowest-unoccupied top right; minority-spin highest-occupied bottom left; majority-spin highest-occupied bottom right. The results shown here are from the calculation with the modified PBE0 functional with 40 % of Fock exchange.

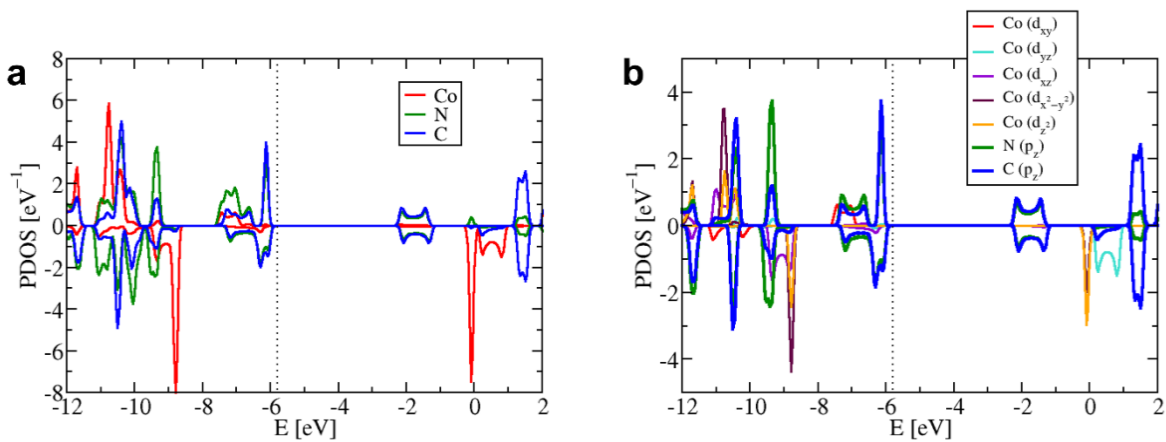

**Figure S9.** Species- and orbital-projected densities of states in the **CoQDI** chain in the ferromagnetic **high-spin** state. (a) Species-projected DOS: Red lines chromium atoms, dark green lines nitrogen atoms, and blue lines carbon atoms. (b) Orbital-projected DOS: red, turquois, violet, maroon, and orange lines stand for $d_{xy}$, $d_{yz}$, $d_{xz}$, $d_{x^2-y^2}$, and $d_{z^2-r^2/3}$ orbitals of Co, respectively, while dark green and blue stand for $p_z$ orbitals of nitrogen and carbon atoms, respectively. Positive values represent majority spin, negative values minority spin. The results shown here are from the calculation with the modified PBE0 functional with 40 % of Fock exchange.

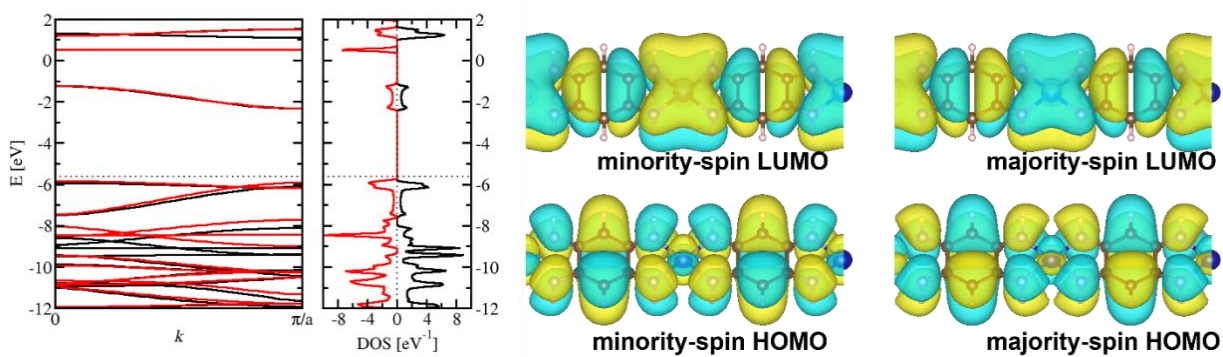

**Figure S10.** Band structure of the **CoQDI** chain in the ferromagnetic **low-spin** state (leftmost panel), the corresponding density of states (next panel to the right), and frontier orbitals (more to the right). Black solid lines represent the majority spin, red solid lines minority spin states. Black dotted line denotes the Fermi level. Depicted orbitals: Minority-spin lowest-unoccupied top left; majority spin lowest-unoccupied top right; minority-spin highest-occupied bottom left; majority-spin highest-occupied bottom right. The results shown here are from the calculation with the modified PBE0 functional with 40 % of Fock exchange.

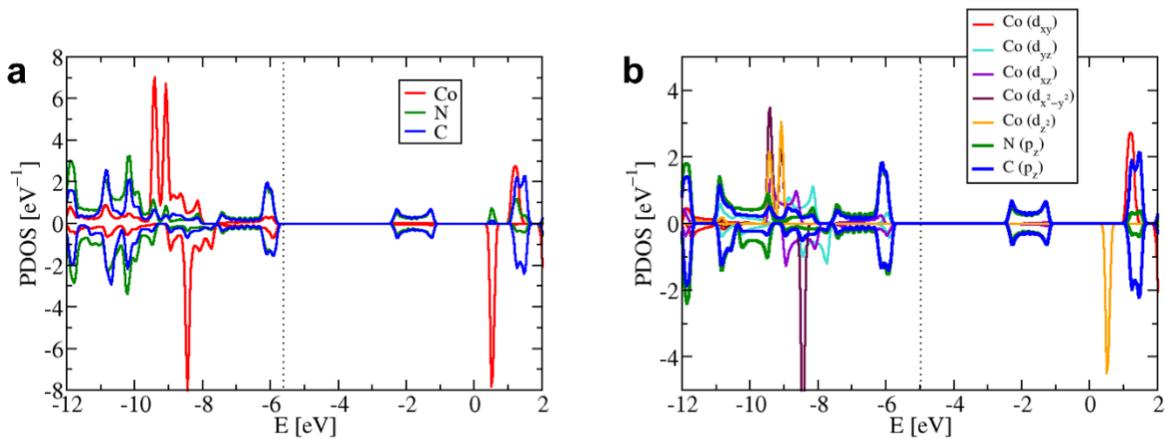

**Figure S11.** Species- and orbital-projected densities of states in the **CoQDI** chain in the ferromagnetic **low-spin** state. (a) Species-projected DOS: Red lines chromium atoms, dark green lines nitrogen atoms, and blue lines carbon atoms. (b) Orbital-projected DOS: red, turquois, violet, maroon, and orange lines stand for $d_{xy}$, $d_{yz}$, $d_{xz}$, $d_{x^2-y^2}$, and $d_{z^2-r^2/3}$ orbitals of Co, respectively, while dark green and blue stand for $p_z$ orbitals of nitrogen and carbon atoms, respectively. Positive values represent majority spin, negative values minority spin. The results shown here are from the calculation with the modified PBE0 functional with 40 % of Fock exchange.

**Table S7. Magnetic properties of CrQDI and CoQDI calculated with DFT, using a hybrid functional with 40 % of Fock exchange.** Low-spin–to–high-spin energy difference $\Delta E = (E_{LS} - E_{HS})/2$ (the denominator of 2 appears because AFM unit cells with two metal atoms are used); exchange coupling $J = E_{AFM} - E_{FM}/(2 \times S^2)$ with $S = m_s / 2$. No low-spin solution was found for Cr.

| Metal atom | Metal−Nitrogen bond length (Å) | $\Delta E$ (meV) | $J$ (meV)* | $m_s$ ($\mu_B$) |
|---|---|---|---|---|
| Co (LS) | 1.91 | +22.97 | −1.50 (AFM) | 1 |
| Co (HS) | 2.02 | 0 | −1.61 (AFM) | 3 |
| Cr | 2.04 | - | −6.66 (AFM) | 4 |

## Additional STM spectroscopy data and fit of the Kondo resonance

Fano resonance fits (Figure S13a) were performed by fitting the data points with a Fano line shape[1,2]:

$$\rho(E) \propto \frac{(q + \epsilon)^2}{1 + \epsilon^2},$$

convolved with Fermi-Dirac distribution for $T = 1.2$ K (temperature of the microscope head), where $\epsilon$ is the normalized energy:

$$\epsilon = \frac{E - E_k}{\Gamma_{Fano}}$$

The Kondo temperature was extracted from the Fano resonance half width at half maximum $\Gamma_{Fano}$ using equation[2]:

$$2\,\Gamma_{Fano} = 2\sqrt{(\pi k_B T)^2 + 2(k_B T_K{}^2)}$$

For the temperature of the system $T = 1.2$ K, the single point estimate of the Kondo temperature from the Fano resonance fit is $T_K = 34.0$ K.

For comparison (Figure S13b), we have also fitted the data with the Frota function[3,4]:

$$\rho_F(\epsilon) \propto \mathrm{Im}\left[-i\sqrt{\frac{i\Gamma_{Frota}}{i\Gamma_{Frota} + \epsilon}}\right].$$

In this case, the Kondo temperature was extracted using the procedure described in ref. 5 by correcting the extracted half-width at half-maximum $\Gamma_{Frota}$ by a simple quadratic approximation:

$$\tilde{\Gamma} = \sqrt{\Gamma^2 - \left(\frac{1}{2} 3.5 k_B T\right)^2},$$

where the second term accounts for the thermal broadening at system temperature (T = 1.2 K). The corrected value is consequently used to estimate the $T_K$, using the same equation as in the case of Fano resonance. For the Frota fit, we have obtained a single point estimate of the Kondo temperature $T_K = 35.7$ K.

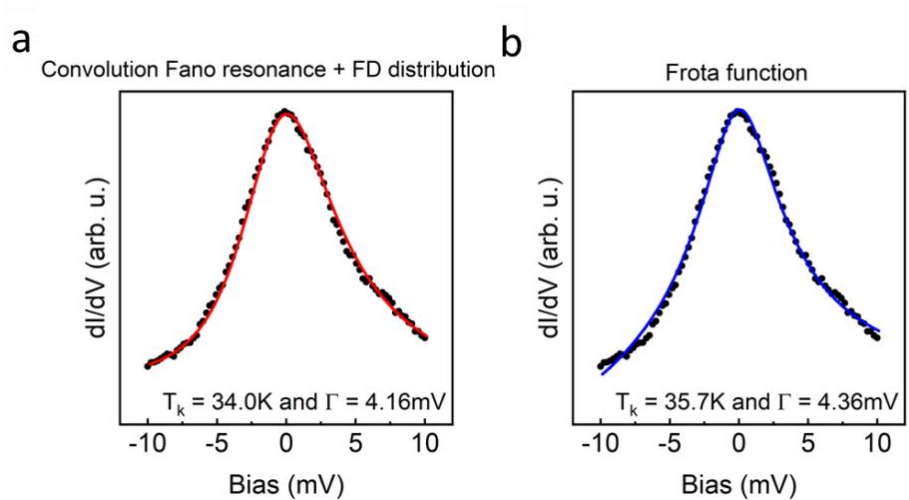

**Figure S12.** Fit of the Kondo peak measured on Cr site of CrQDI chains, see Fig. 5. The estimates of Kondo temperature are based on fits of a) thermally broadened Fano resonance and b) Frota function. For the high Kondo temperatures (~35 K; ~ 4 meV) neglecting the very small broadening due to the lock-in amplifier (0.5 mV rms) is justified.

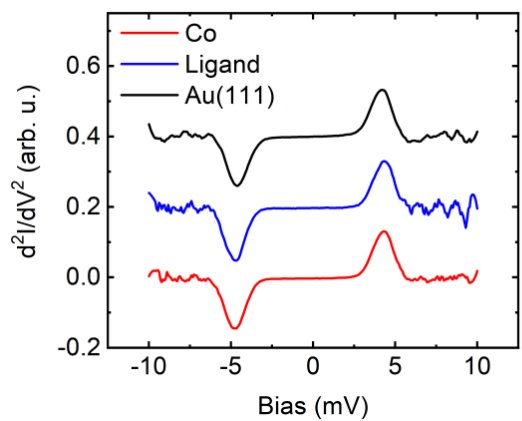

**Figure S13**. IETS spectra comparison of NiCp2 tip taken on different sites in the CoQDI experiment. Red, blue and black curves correspond to spectra taken on Co, ligand and gold substrate sites, respectively. Curves smoothed to reduce the noise. Tip-sample distance for all curves corresponds to STM set point of ~3.2 nA at 10 mV.

# Simulations of IETS in the Hubbard model

**Table S7. Parameters of the individual 3d shells constituting the Hubbard model.** Crystal-field parameters $B_{kq}$ in the Wybourne notation, strength of the spin-orbit coupling $\zeta$, energy of the 3d level $\varepsilon$, and Slater integrals $F_k$. All values are shown in eV.

|          | $Cr^{2+}$ ($C_{2v}$) | $Co^{2+}$ ($C_{2v}$) | $Ni^{2+}$ ($D_{5d}$) |
|----------|---------|---------|---------|
| $B_{20}$ | −2.93   | −5.72   | −0.40   |
| $B_{22}$ | −1.67   | 1.70    | —       |
| $B_{40}$ | 1.24    | 1.15    | −8.28   |
| $B_{42}$ | 0.20    | −0.26   | —       |
| $B_{44}$ | −2.91   | −1.42   | —       |
| $\zeta$  | 0.030   | 0.068   | 0.074   |
| $\varepsilon$ | −25.00 | −50.00 | −58.60 |
| $F_0$    | 8.00    | 8.00    | 8.00    |
| $F_2$    | 7.70    | 8.42    | 9.30    |
| $F_4$    | 4.79    | 5.23    | 5.77    |

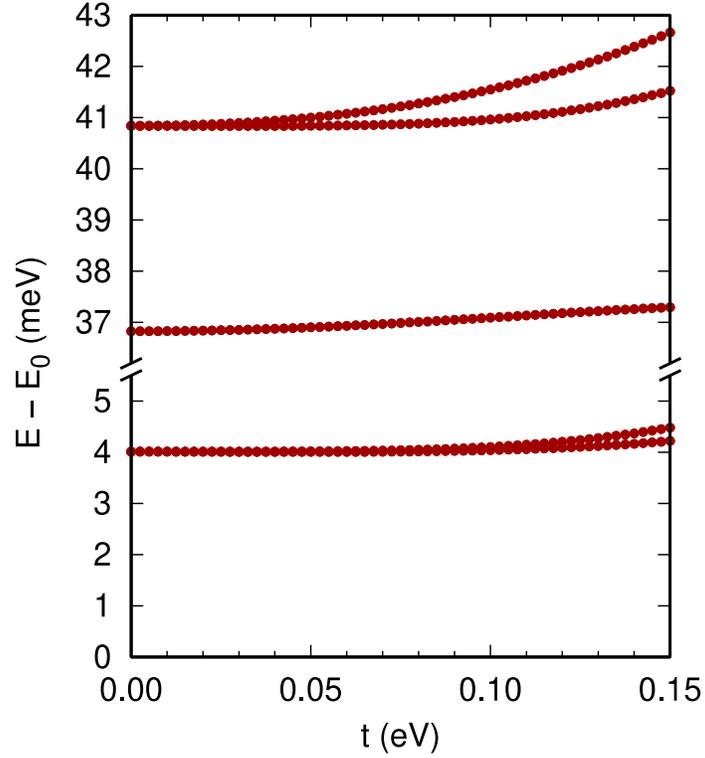

**Figure S14. Hubbard model for NiCp$_2$ tip and CoQDI.** Evolution of the excitation energies measured with respect to the ground-state energy $E_0$. The excitation at 4 meV corresponds to a spin flip at the Ni site, the excitation at 37 meV corresponds to a spin flip at the Co site and the excitation at 4 + 37 = 41 meV corresponds to spin flips at both sites.

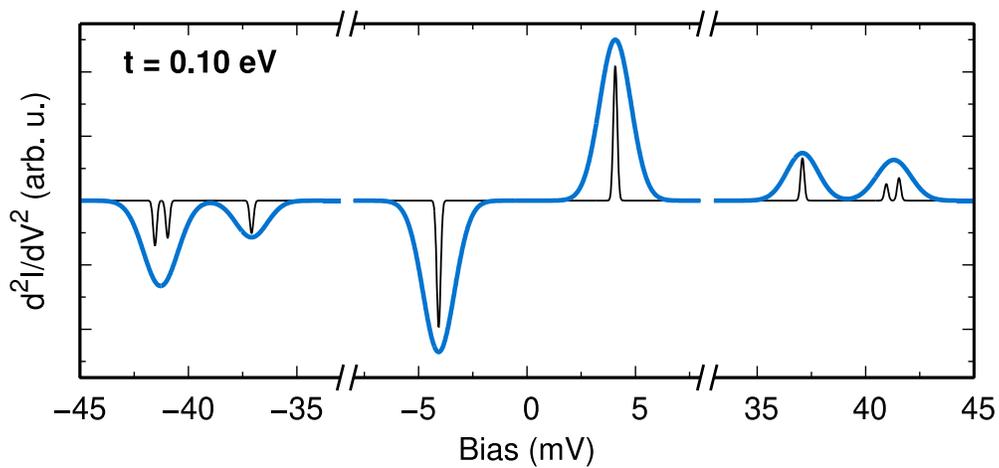

**Figure S15. Theoretical d$^2$I/dV$^2$ spectrum of NiCp$_2$ tip and CoQDI** plotted over a larger bias range than in Figure 7e to capture the peaks involving spin flips at the Co site.